\documentclass{nature}

\usepackage[T1]{fontenc}
\usepackage{textcomp}
\input{glyphtounicode}
\PassOptionsToPackage{hyphens}{url}
\usepackage{doi}
\usepackage{hyperref}
\usepackage{amssymb}
\usepackage{amsmath}
\usepackage{fix-cm}
\usepackage{graphicx}
\usepackage{xspace}
\usepackage{multirow}
\usepackage[usenames, dvipsnames]{color}
\usepackage[normalem]{ulem}
\usepackage{soul}
\usepackage{subfigure} 
\usepackage{graphicx}
\usepackage{gensymb}
\usepackage[mathlines]{lineno}
\usepackage{subfigure}
\usepackage[font=footnotesize]{caption}
\DeclareCaptionLabelSeparator{journalbar}{~|~}
\usepackage{floatrow}
\DeclareFloatFont{tiny}{\scriptsize}

\title{An intertwined short and long $\gamma$-ray burst with a 4-minute separation}

\begin{document}

\author{
Liang~Li$^{\ref{IFPQT},\ref{NBU},\ref{BINGO},\ref{ZKLEU}*}$, 
Yu~Wang$^{\ref{ICRA},\ref{ICRANet},\ref{INAF},\ref{MGC}*}$,
Bing~Zhang$^{\ref{HKU1},\ref{HKU2}*}$,
Ye~Li$^{\ref{PMO}}$, 
Shu-Rui~Zhang$^{\ref{ICRA},\ref{ICRANet},\ref{MGC}}$,
Jochen~Greiner$^{\ref{MPE}}$,
Zhi-Ping~Jin$^{\ref{PMO}}$,
Jin-Jun~Geng$^{\ref{PMO}}$,
Hou-Jun~L\"{u}$^{\ref{GKLRA}}$,
Asaf~Pe{\textquoteright}er$^{\ref{DPBI}}$,
Maria~Giovanna~Dainotti$^{\ref{NAOJ}}$,
Tong~Liu$^{\ref{DAXU}}$,
Yi-Zhong~Fan$^{\ref{PMO}}$,
Yong-Feng~Huang$^{\ref{NU},\ref{KLMAA}}$,
Zi-Gao~Dai$^{\ref{USTC}}$,
Melin~Kole$^{\ref{UNH},\ref{UG}}$, 
Wei-Hua~Lei$^{\ref{DASP}}$,
Ye-Fei~Yuan$^{\ref{USTC}}$,
Shuang-Nan~Zhang$^{\ref{IHEP}}$, 
Felix~Ryde$^{\ref{KTH}}$,
She-Sheng~Xue$^{\ref{ICRA},\ref{ICRANet}}$ \&
Rong-Gen~Cai$^{\ref{IFPQT},\ref{ZKLEU}*}$
}

\maketitle

\noindent{\small Nature Astronomy\quad Article\quad\url{https://doi.org/10.1038/s41550-026-03007-0}}\\
\noindent{\small Received: 13 May 2026\quad Accepted: 22 September 2026\quad Published online: 23 October 2026}

\begin{affiliations}
\small 
\item Institute of Fundamental Physics and Quantum Technology, Ningbo University, Ningbo, China\label{IFPQT}.
\item Department of Physics, School of Physical Science and Technology, Ningbo University, Ningbo, China\label{NBU}.
\item BINGO Center, Ningbo University, Ningbo, China\label{BINGO}.
\item Zhejiang Key Laboratory of Extreme Universe, Ningbo University, Ningbo, China\label{ZKLEU}.
\item Dipartimento di Fisica and ICRA, Sapienza Universita di Roma, Rome, Italy\label{ICRA}.
\item ICRANet, Pescara, Italy\label{ICRANet}.
\item INAF -- Osservatorio Astronomico d'Abruzzo, Teramo, Italy\label{INAF}.
\item Marcel Grossmann Center, Miami, FL, USA\label{MGC}.
\item The Hong Kong Institute for Astronomy and Astrophysics, The University of Hong Kong, Hong Kong, China\label{HKU1}.
\item Department of Physics, The University of Hong Kong, Hong Kong, China\label{HKU2}.
\item Purple Mountain Observatory, Chinese Academy of Sciences, Nanjing, China\label{PMO}.
\item Max-Planck Institut für extraterrestrische Physik, Garching, Germany\label{MPE}. 
\item Guangxi Key Laboratory for Relativistic Astrophysics, School of Physical Science and Technology, Guangxi University, Nanning, China\label{GKLRA}.
\item Department of Physics, Bar-Ilan University, Ramat-Gan, Israel\label{DPBI}.
\item Division of Science, National Astronomical Observatory of Japan, Osawa, Tokyo, Japan\label{NAOJ}.
\item Department of Astronomy, Xiamen University, Xiamen, China\label{DAXU}.
\item School of Astronomy and Space Science, Nanjing University, Nanjing, China\label{NU}.
\item Key Laboratory of Modern Astronomy and Astrophysics (Nanjing University), Ministry of Education, Nanjing, China\label{KLMAA}.
\item Department of Astronomy, University of Science and Technology of China, Hefei, China\label{USTC}.
\item University of New Hampshire, Space Science Center, University of New Hampshire, Durham, NH, USA\label{UNH}.
\item Department of Nuclear and Particle Physics, University of Geneva, Geneva, Switzerland\label{UG}.
\item Department of Astronomy, School of Physics, Huazhong University of Science and Technology, Wuhan, China\label{DASP}.
\item Key Laboratory of Particle Astrophysics, Institute of High Energy Physics, Chinese Academy of Sciences, Beijing, China\label{IHEP}.
\item Department of Physics, KTH Royal Institute of Technology, and The Oskar Klein Centre, Stockholm, Sweden\label{KTH}.
\end{affiliations}
\noindent e-mail: liliang@nbu.edu.cn; yu.wang@icranet.org; zhang@physics.unlv.edu; cairg@itp.ac.cn

\begin{abstract} 

Gamma-ray bursts (GRBs), among the most energetic transients in the Universe, are traditionally classified into long-duration (lasting more than 2 seconds) and short-duration (lasting less than 2 seconds) events. Long-duration GRBs are typically associated with the core collapse of massive stars (type II), whereas short-duration GRBs originate from the merger of compact binary systems (type I). These different progenitor channels are reflected in distinct observational signatures that provide key diagnostics for GRB classification. Here we report a peculiar event, GRB 160425A, comprising two sub-bursts separated by 4 minutes: a short-duration burst ($G_1$) and a long-duration burst ($G_2$). Nearly all standard prompt-emission diagnostics, including pulse morphology, duration, hardness ratio, minimum variability timescale, spectral properties and established empirical correlations, consistently categorize $G_1$ as a short-like GRB (type I, merger origin) and $G_2$ as a long-like (type II, collapsar origin) GRB. The coexistence of merger and collapsar signatures within a single event challenges existing progenitor frameworks and calls for a re-evaluation of GRB classification schemes and progenitor scenarios. 

{\bf  }

\end{abstract}

{\bf Observations of GRB 160425A}

\begin{figure}[ht!]
\centering
\includegraphics[width=\textwidth,height=0.62\textheight,keepaspectratio]{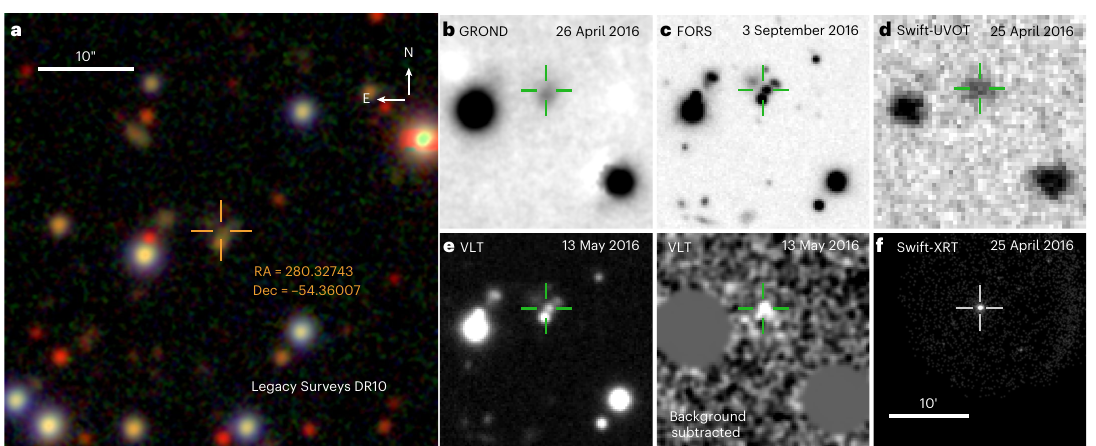}
\caption{{\bf Images of GRB 160425A and its host galaxy.} {\bf a}, The survey data from Dark Energy Spectroscopic Instrument (DESI) Legacy Surveys Data Release 10 (DR10)\cite{Dey2019AJ}, where the region with a knot structure at the centre represents the host galaxy of GRB 160425A. {\bf b-e}, Observations of the central celestial region in {\bf a}, captured by different optical telescopes at various times. {\bf b}, Observations by the GROND telescope on 26 April. {\bf c}, Observations by the FORS telescope of the VLT on 3 September, revealing three diffuse knots. {\bf d}, Data from \emph{Swift}-UVOT on 25 April. {\bf e}, Left: observations by the VLT on 13 May. Right: the same region with the background galaxies subtracted, revealing the optical image of the GRB. {\bf f}, Observations of soft X-rays by \emph{Swift}-XRT on 25 April, showcasing a larger field of view compared with the images mentioned above. In all panels, the crosshairs mark the position of the optical afterglow; their arm lengths are included only as visual guides and do not represent positional uncertainties.\label{fig:HG_image}}
\end{figure}

\begin{figure}[ht!]
\centering
\includegraphics[width=\textwidth,height=0.62\textheight,keepaspectratio]{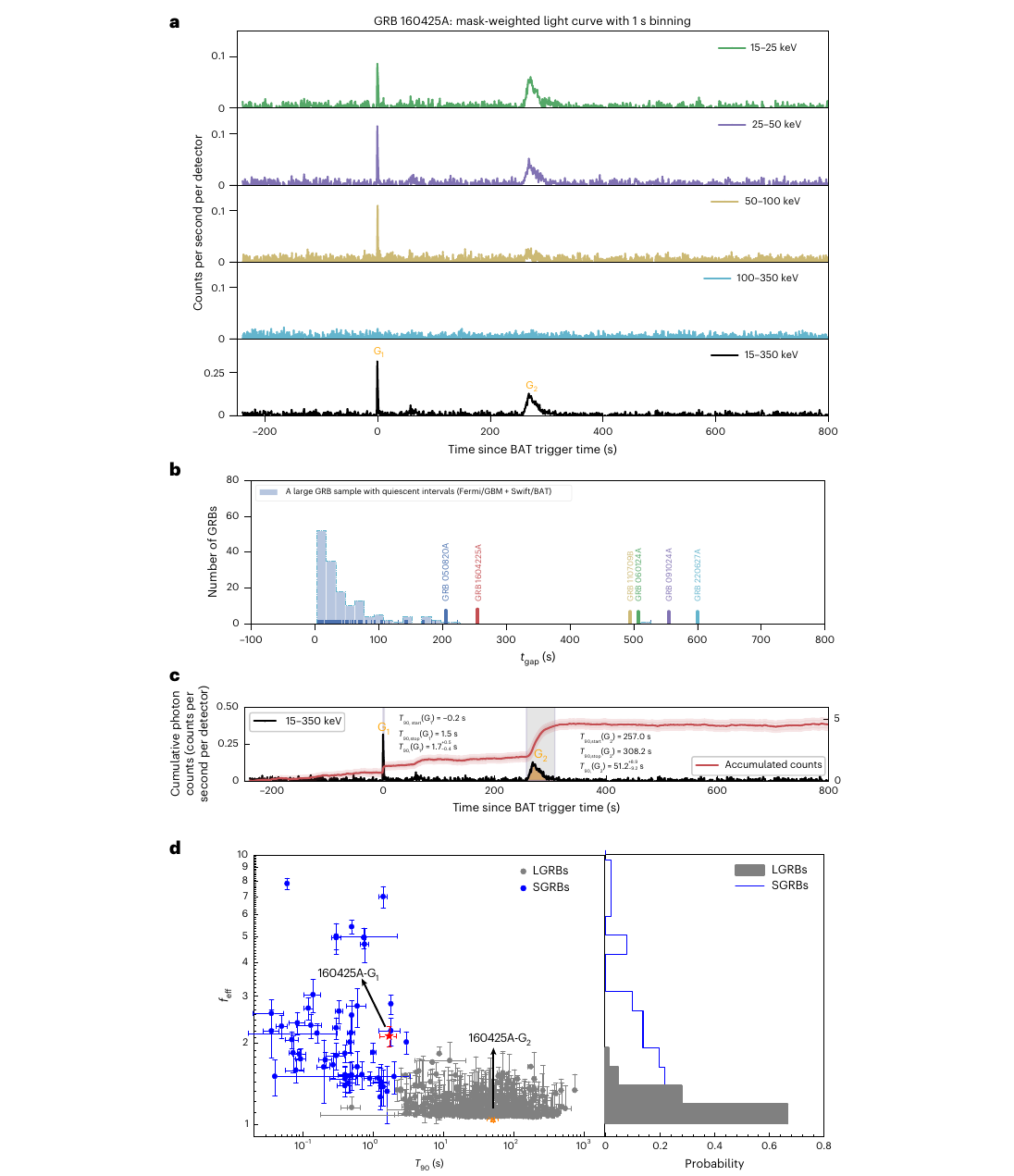}
\caption{{\bf Prompt $\gamma$-ray emission phase of GRB 160425A.} {\bf a}, The background-subtracted light curves of GRB 160425A in distinct energy bands (green, 15-25 keV; magenta, 25-50 keV; yellow, 50-100 keV; cyan, 100-350 keV; black, 15-350 keV), derived from \emph{Swift}/BAT data. {\bf b}, GRB 160425A in the distribution of quiescent period $t_{\rm gap}$ compared with other GRBs with long quiescent intervals. {\bf c}, $T_{90}$ analysis of the two individual bursts of GRB 160425A. The black curve (left) shows the \emph{Swift}-BAT light curve over the full 15-350 keV energy range and the red curve (right) shows the measured cumulative counts. The pink shaded band represents the propagated 1$\sigma$ uncertainty of the cumulative counts, derived from the per-bin 1$\sigma$ statistical uncertainties of the \emph{Swift}-BAT photon counts. {\bf d}, Left: $T_{90}$-$f$($f_{\rm eff}$) scatter plot for SGRBs (blue circles) and LGRBs (grey circles), where $f$ represents the ratio between the peak flux and the average background flux of a GRB (Methods). $f_{\rm eff}$ is the effective $f$ parameter, indicating how a LGRB can be disguised as an SGRB by arbitrarily lowering its flux level. The data points represent individual measured values, and the horizontal and vertical error bars represent the corresponding 1$\sigma$ measurement uncertainties. The two distinct bursts of GRB 160425A are highlighted by the red ($G_1$) and orange ($G_2$) solid stars, respectively. Right: probability distribution of $f_{\rm eff}$ for SGRBs (blue line) and LGRBs (grey-filled histogram).\label{fig:lcs_PE}}
\end{figure}

On 25 April 2016, at 23:26:11 UT, GRB 160425A alerted the Burst Alert Telescope (BAT)\cite{2016GCN.19343....1K} onboard the Neil Gehrels {\it Swift} Observatory, Mobile Astronomical System of Telescope-Robots (MASTER) II, Max Planck Gesellschaft (MPG) (2.2 m), Very Large Telescope (VLT), MASTER-South African Astronomical Observatory (SAAO) and Skynet Panoramic Robotic Optical Monitoring and Polarimetry Telescopes (PROMPT) (Fig. \ref{fig:HG_image}). The net {\it Swift}/BAT light curve along with its light curve in different energy bands (15-25 keV to 100-350 keV), as detected by BAT, is shown in Fig. \ref{fig:lcs_PE}a. Long-term follow-up observations detected the optical counterpart\cite{2016GCN.19348....1M,2016GCN.19343....1K,2016GCN.19349....1B,2016GCN.19354....1C,2016GCN.19357....1S} and identified its host galaxy and redshift at $z=0.555$ (ref.~\citen{Tanvir2016}). A summary of the timing, spectral properties and host-galaxy information of the burst is listed in Table \ref{tab:Global}.

{\bf Direct prompt-emission observables} 

One distinct feature of GRB 160425A is that it comprises two sub-bursts $G_1$ and $G_2$, separated by a quiescent interval exceeding 250 s during which no significant emission is detected (Fig. \ref{fig:lcs_PE}b). The two sub-bursts appear to be independent, providing a unique opportunity to study GRB classification (see Methods for details). The first sub-burst, $G_1$, is particularly interesting owing to its short duration ($\sim$ 1.7~s) and a sharply peaked pulse profile (Fig. \ref{fig:lcs_PE}c and Methods), morphologically resembling a classical short-duration GRB (SGRB), which raises the question of whether it is related to the second sub-burst, $G_2$. We further measure the `amplitude parameter' $f$ of $G_{1}$, defined as the ratio between the peak flux and the average background flux of a GRB (Methods). We obtain $f=2.13\pm0.18$ for $G_{1}$, which is higher than the low-$f$ range usually associated with disguised short GRBs, indicating that $G_1$ favours a genuinely short-duration burst and disfavours a tip-of-the-iceberg interpretation (Fig. \ref{fig:lcs_PE}d). The second sub-burst, $G_{2}$, has $T_{90}\approx 51.2$~s (Fig. \ref{fig:lcs_PE}c and Methods), consistent with a canonical long-duration GRB (LGRB). The timing and spectral properties of $G_1$ and $G_2$  are summarized in Table \ref{tab:IndBurst}.

The positions of $G_1$ and $G_2$ are determined independently and consistent with the same BAT location within positional uncertainties, further supported by the common afterglow/host galaxy association (Methods), suggesting that both sub-bursts probably originate from the same astrophysical system. The host association itself is based on the fading X-ray/optical afterglow and its arcsecond-level position on the $z=0.555$ host galaxy.

\begin{figure}[ht!]
\centering
\includegraphics[width=\textwidth,height=0.62\textheight,keepaspectratio]{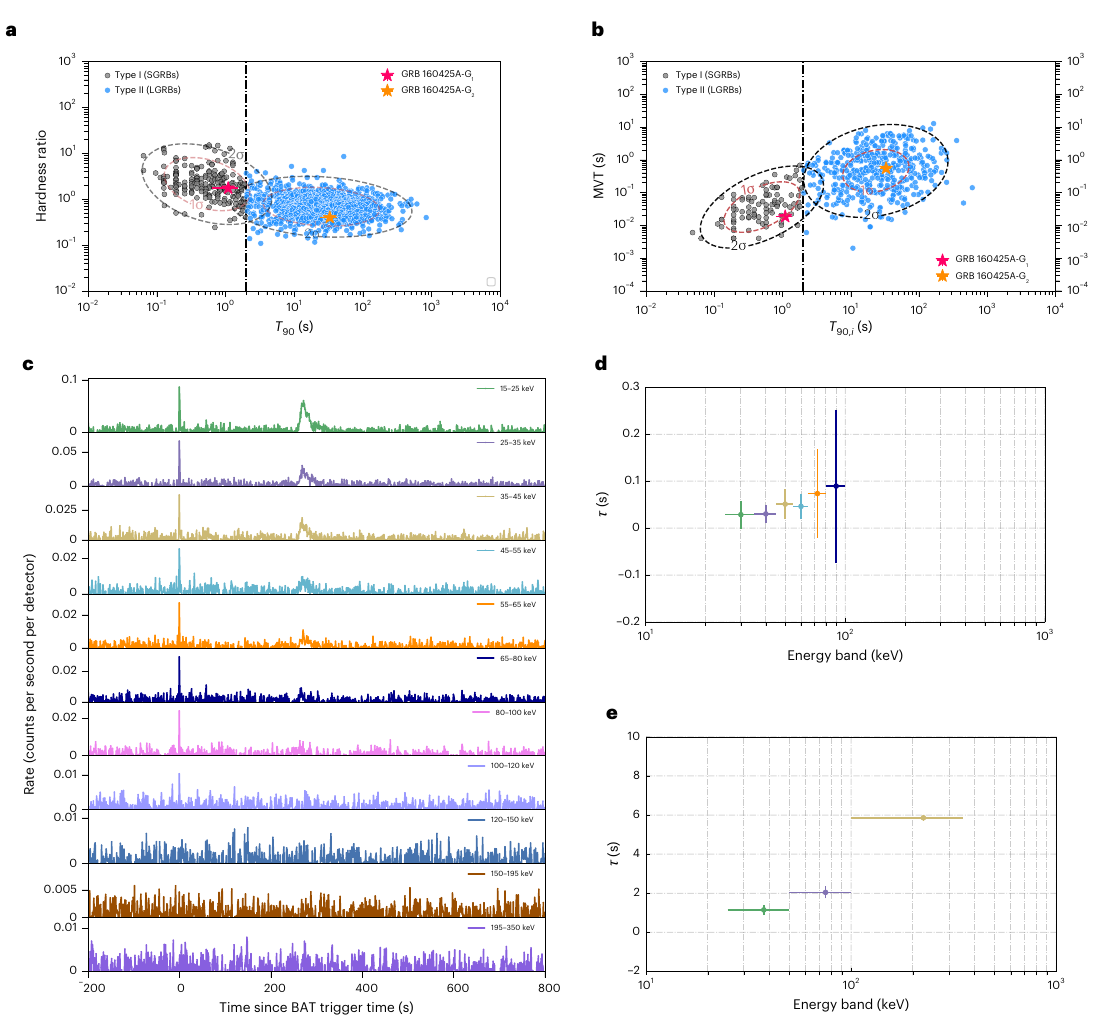}
\caption{{\bf GRB classification scheme.} {\bf a,b}, The traditional GRB classification is illustrated based on the duration-hardness ratio diagram ({\bf a}) and the duration-MVT diagram ({\bf b}). The \emph{Swift} GRB sample is represented by solid grey and cyan points for the short-duration and long-duration burst populations, respectively. The two individual bursts of GRB 160425A are highlighted by red triangular and orange stars. Elliptical dotted lines in different colours indicate the 1$\sigma$ and 2$\sigma$ regions of the bivariate normal distributions for the short-duration and long-duration burst populations. The traditional separation line ($T_{90}$ = 2 s) between SGRBs and LGRBs is indicated by a black dashed vertical line. Duration is plotted in the observed frame for a and the source frame for {\bf b}. {\bf c-e}, Spectral lag analysis. c, Background-subtracted \emph{Swift} light curves of GRB 160425A in different energy bands, used to calculate spectral lags. {\bf d,e}, Energy-dependent spectral lag between the lowest energy band (15-25 keV) and the higher-energy band for the short-duration burst ($G_1$) ({\bf d}) and the long-duration burst ($G_2$) ({\bf e}) of GRB 160425A.\label{fig:Classification}}
\end{figure}

To test whether the first sub-burst, $G_1$, is related to the second sub-burst, $G_2$, we compared both sub-bursts with traditional GRB classification schemes (Fig. \ref{fig:Classification}, Extended Data Fig.~1 and Methods). Figure \ref{fig:Classification}a shows the standard long/short duration–hardness ratio plane, where the hardness ratio is defined as the ratio of the counts in the 25-50 keV band to those in the 15-25 keV band, accumulated over the $T_{90}$ duration of each sub-burst, the time interval containing 90\% of the burst fluence (Methods). One can see that $G_{1}$ occupies the short-hard region of this diagram, while $G_{2}$ falls squarely within the long-soft distribution. 

The minimum variability timescale (MVT, $\Delta t_{\rm min}$) is the shortest observable timescale over which significant flux variations are observed in a GRB and provides an additional discriminator between GRB classes. On the basis of a Fermi GRB sample\cite{Golkhou2015}, the median MVT is 18~ms for SGRBs and 134~ms for LGRBs (Methods). We measure $\Delta t_{\rm min}=18.5 \pm 3.1$~ms for $G_1$ and $\Delta t_{\rm min}=550\pm32$~ms for $G_2$ (Fig. \ref{fig:Classification}b and Methods). These results suggest that $G_1$ exhibits characteristics consistent with SGRBs (type I), while $G_2$ aligns with the properties of LGRBs (type II).

Observations show that SGRBs typically exhibit a negligible `spectral lag' whereas LGRBs are associated with a non-negligible spectral lag\cite{Yi2006,ZhangZhibin2006,Bernardini2015}. We measure the 15-25~keV to 25-50~keV spectral lag for each sub-burst and find 38$\pm$39~ms for $G_{1}$ and 1,152 $\pm$264~ms for $G_{2}$ (Fig. \ref{fig:Classification}c-e and Methods). The lag of $G_1$ is consistent with zero, as expected for a type ~I burst, while the highly significant lag of $G_2$ is characteristic of type ~II GRBs. Similar results are also found in the luminosity–lag ($L_{\rm p,iso}-\tau$) plane. Plotting the isotropic peak luminosity of each sub-burst on the Norris-relation diagram\cite{Norris2000} (Extended Data Fig.~1a), we find that $G_{2}$ follows the relation defined by long-duration GRBs, whereas $G_{1}$ deviates markedly from it, as expected for short-duration GRBs.

{\bf Derived prompt-emission diagnostics}

We perform the spectral analysis of $G_{1}$ and $G_{2}$ (see \emph{Swift} data analysis in Methods and Extended Data Fig.~2). No contemporaneous Fermi Gamma-ray Burst Monitor (GBM) data are available for this burst, so the spectral constraints are based on \emph{Swift}/BAT data alone (Extended Data Fig.~3 and Methods). The spectral peak energy is estimated as $E_{\rm p}\gtrsim 500$ keV for $G_{1}$ (see Methods for details) and $E_{\rm p}=57.6\pm2.7$ keV for $G_{2}$ (Methods). The corresponding observed fluences are (8.6$\pm$1.1)$\times$10$^{-7}$ erg cm$^{-2}$ for $G_{1}$ and (2.5$\pm$0.3)$\times$10$^{-6}$ erg cm$^{-2}$ for $G_{2}$. At the host-galaxy redshift $z$=0.555 (ref.~\citen{Tanvir2016}), the isotropic $\gamma$-ray energy ($E_{\gamma,\rm iso}$) of $G_{1}$ can be estimated as $E_{\gamma,\rm iso}$ $\approx$ (7.2$\pm$0.9) $\times$ 10$^{50}$ erg, and that of $G_{2}$ as $E_{\gamma,\rm iso}$ $\approx$ (2.1 $\pm$ 0.3) $\times$ 10$^{51}$ erg. We over-plotted the $E_{\rm p,z}$-$E_{\gamma,\rm iso}$ diagram\cite{Amati2002,Zhang2009b,Qin2013}, and find that $G_{1}$ falls within the short-duration (type ~I) locus while $G_{2}$ lies in the long-duration (type ~II) locus (Extended Data Fig.~1b).

The peculiarity of two sub-bursts observed in GRB 160425A is also reflected in its value of the energy-hardness-duration (EHD) parameter\cite{Minaev2020}, EHD=$E_{\rm p,i,2} E^{-0.4}_{\rm iso,51}T^{-0.5}_{90,i}$, where $E_{\rm p,i,2}$ is the rest-frame peak energy $E_{\rm p,i}$ in units of 100 keV, $E_{\rm iso,51}$ is the isotropic equivalent energy $E_{\gamma, \rm iso}$ in units of $10^{51}$ erg, and $T_{90,i}$ is the rest-frame $T_{90}$, expressed in seconds. Reference \citen{Minaev2020} demonstrated that EHD provides the most reliable parameter for the blind type I and type II classification, with the two populations divided at EHD=2.6, with type I GRBs clustering in the high-EHD region and type II GRBs in the low-EHD region (Methods). We obtain EHD=9.5 for $G_{1}$ and EHD=0.01 for $G_{2}$, placing $G_{1}$ firmly in the type ~I region and $G_{2}$ firmly in the type ~II region of the EHD-$T_{90,i}$ diagram (Extended Data Fig.~1c).

We extend the classification analysis to the phenomenological parameter $\varepsilon \equiv E_{\gamma,\rm iso,52}/E^{5/3}_{\rm p,z,2}$, which has been found to be successful in classifying type I versus type II GRBs\cite{Lv2010}, where $E_{\gamma,\rm iso,52}$ is the isotropic equivalent energy in units of $10^{52}$ erg and $E_{\rm p,z,2}$ is the cosmological rest-frame spectral peak energy in units of 100~keV. The two GRB classes are nicely clustered in high-$\varepsilon$ ($\varepsilon>$0.03) and low-$\varepsilon$ ($\varepsilon<$0.03) regions at the separation line by $\varepsilon \approx$ 0.03 using a GRB sample with known redshift\cite{Lv2010} (Methods). For GRB~160425A, we find that $\varepsilon$ places $G_{1}$ in the low-$\varepsilon$ (type I) region and $G_{2}$ in the high-$\varepsilon$ (type II) region of the $\log\varepsilon$-$\log T_{90,z}$ plane (Extended Data Fig.~1d).

{\bf Uniqueness of GRB 160425A in the peculiar GRB population}

GRB~160425A is uniquely distinguished from all previously reported peculiar GRBs by the simultaneous presence of three properties: two genuine prompt-like emission episodes, an ${\sim}256$-s background-consistent quiescent gap, and opposite empirical classifications across the gap ($G_1$ short-like; $G_2$ long-like). Long-duration merger-like events such as GRB 060614\cite{Fynbo2006,GalYam2006,Gehrels2006}, GRB 211211A\cite{Rastinejad2022,Troja2022,Yang2022}  and GRB 230307A\cite{Levan2024,Yang2024Y}, short-duration collapsar-like events such as GRB 200826A\cite{ZhangBB2021}, high-redshift rest-frame short events such as GRB 080913 and GRB 090423\cite{Zhang2009b}, and boundary cases such as GRB 180418A\cite{Becerra2019} mainly expose the limitation of total-duration classification. They do not show two well-separated prompt-like emission episodes with opposite empirical classifications. GRB 160425A also differs greatly from previously reported quiescent multi-episode GRBs. Precursor ($G_1$)-main ($G_2$) emission bursts\cite{Li2026b} open with a weak early episode before the dominant long-duration prompt emission (for example, GRB 060124); main ($G_1$)-extended ($G_2$) emission events\cite{Li2026d} close with a softer, fainter tail (for example, GRB 100704A); and double-trigger events\cite{Li2026c} comprise two sub-bursts separated by a quiescent interval long enough to produce independent instrument triggers (for example, GRB 110709B). In all these cases, the temporally separated episodes share a common long-like classification. By contrast, SGRBs ($G_1$) with a fainter extended-emission tail ($G_2$)\cite{Norris2006} (for example, GRB 050709), late X-ray flares ($G_2$)\cite{Falcone2006,Dai2006} (for example, GRB 050724)  and afterglow-onset components ($G_2$)\cite{Hakkila2004} (for example, GRB 960530) are not ordinary long-like prompt-emission episodes (Table \ref{tab:GRBtype} and Methods).

{\bf Afterglow, supernova  and host-galaxy constraints}

Long-term follow-up multiwavelength afterglow observations of the burst were carried out by {\it Swift}\cite{2016GCN.19348....1M} and several ground-based optical telescopes\cite{2016GCN.19343....1K,2016GCN.19349....1B,2016GCN.19354....1C,2016GCN.19357....1S} (Extended Data Tables~1 and 2 and Methods). In general, SGRBs have lower luminosity than typical LGRBs in X-ray afterglows as observed with the \emph{Swift} X-ray Telescope (XRT)\cite{Li2015,Li2018b,Ruffini2018,Zhang2018} (Methods). The luminosity of the X-ray afterglow emission is shown as a function of rest-frame time (Extended Data Fig.~4a). Coincidentally, GRB 160425A resides in the transitional region between short-duration and long-duration bursts, situated at the lower boundary of the long-burst distribution and the upper boundary of the short-burst distribution.

GRB 160425A differs from the typical GRB population by having a higher electron-energy fraction ($\epsilon_e\approx0.62$ versus $\epsilon_e\approx0.2$-$0.3$), a higher Lorentz factor ($\Gamma_0\approx950$ versus $\Gamma_0\approx250$ for LGRBs and $\Gamma_0\approx420$ for SGRBs), and a lower circumburst density ($n_0\approx1.6\times10^{-3}\,\mathrm{cm^{-3}}$ versus $n_0\approx0.3\,\mathrm{cm^{-3}}$). The accuracy of these fitted value is limited by the degeneracies among the forward-shock parameters (Extended Data Fig.~5  and Methods).

{\em Swift} optical U-band data were adopted to identify a possible supernova component. After subtracting the host galaxy's contribution, the U-band flux at time $\sim 10^6$~s is faint and highly uncertain, making it insufficient to confirm the existence of a supernova component. This result suggests either no supernova component or a very weak one compared with SN 1998bw. This is consistent with the I-band data from Cerro Tololo Inter-American Observatory (CTIO), which found no obvious change of brightness between 7 days and 26 days\cite{2016GCN.19454....1C} (Extended Data Fig.~6 and Methods).

{\bf Why standard interpretations are challenged}

These phenomena have never been observed in a GRB before, challenging the traditional classification scheme of long-duration and short-duration bursts. The peculiar properties of GRB 160425A call for a new interpretation that has not been invoked to explain standard SGRBs or LGRBs. 

There are five possible ways to interpret this peculiar phenomenon, all of which we will show below to be challenging.

The first possibility is that the two bursts in GRB 160425A originate from distinct host galaxies that happen to coincide in space and time. This scenario suggests that two bursts, each from different host galaxies with different redshifts, were observed in the same spatial region within just a 4-minute separation in observed time, with both jets directed towards Earth. The local probability of observing two isolated bursts within a 250-s interval from two different galaxies, with a positional accuracy of 1.5 arcmin using the BAT instrument, is approximately $1.97 \times 10^{-12}$ (Methods). After accounting for the trials factor over the \emph{Swift}/BAT observing period, the corresponding survey-level chance-coincidence probability is of the order $10^{-7}$-$10^{-5}$ (Methods), which remains small.

The second possibility is that the two bursts originated from the same host galaxy but at two different locations, with $G_1$ originating from a compact binary merger and $G_2$ originating from the core collapse of a massive star. The local conditional probability of observing two isolated bursts within 250 s from the same galaxy is approximately $7.9 \times 10^{-13} - 7.9 \times 10^{-12}$ (Methods). Even after allowing for trials over the \emph{Swift}/BAT observing period, two unrelated bursts from the same host within minutes would still be unusual (Methods).

The third possibility is that this is an extreme case of type I GRB, with the initial short burst ($G_{1}$) being followed by a long-duration extended-emission   tail ($G_{2}$) tens (or hundreds) seconds later. However, this interpretation has a few drawbacks. First of all, a prolonged quiescent period of up to 4 minutes after the emission of $G_1$ and a much higher fluence, a comparable peak flux  and the presence of a strong single-peak pulse profile of $G_2$ have never been observed in extended emission, which strongly disfavours this scenario. One may consider $G_{2}$ as a bright X-ray flare following SGRBs\cite{Dai2006,Margutti2011}.  However, an X-ray flare of this brightness has never been observed  either (Extended Data Fig.~4b-d). In fact, $G_2$ aligns with type II rather than type I in the $L_{\rm x,p}-t_{\rm p,z}$ diagram of X-ray flares (Extended Data Fig.~4e,f  and Methods). Moreover, the type I interpretation suffers from additional criticisms. The offset between the GRB location and its host galaxy is calculated to be 1.86 kpc, corresponding to a normalized offset value of 0.7 (Table \ref{tab:Global} and Methods). Statistically, such a small offset is more consistent with type II GRBs than type I GRBs (Extended Data Table~3 and Extended Data Fig.~7).
 
This left us with the fourth interpretation, that is, this is a peculiar type II GRB with $G_2$ being the main burst, but with an initial short-duration precursor emission ($G_1$). Such a scenario could be supported if the putative supernova is confirmed, but it is hardly identified. This scenario also suffers from several drawbacks. Typically, the peak flux of a precursor emission is much lower than that of the main emission of a long burst\cite{Zhang2018NA,Ryde2022}. However, in GRB 160425A, the peak flux of $G_1$ is even higher than that of $G_2$. Moreover, the analysis of the $f$ value supports $G_1$ being a genuine short burst rather than a precursor emission of a long burst. More importantly, previous studies\cite{Burlon2008,Burlon2009,Hu2014,Zhang2018NA,Anand2018,LiXJ2021,Yang2022,DengHY2024} indicate that these emission components--precursors, prompt $\gamma$-ray emission  and extended soft $\gamma$-ray emission--are likely to share a common physical origin, attributed to the repeated activation of the GRB central engine. Consequently, these distinct components within an individual burst typically exhibit similarities in observational properties, such as spectral lags. This is also in contrast with the observations. Physically, a quiescent gap may be produced if the type II GRB central engine is intermittent, as shown in numerical simulations of intermittent jets propagating in a massive-star envelope\cite{geng16}. However, such a process usually produces weak precursors. The origin of an extremely narrow and hard precursor such as $G_1$ is still mysterious.  

{\bf A possible non-standard origin}

A fifth possibility is that GRB 160425A reflects a more complex, non-standard progenitor configuration rather than a conventional type I or type II channel. This possibility could in principle link the two prompt-like episodes within one system, but it remains highly uncertain because the required dynamics, accretion process, energetics  and event rate are not yet well constrained.

{\bf Conclusion and implications}

The direct prompt-emission observables (duration, amplitude parameter, hardness ratio, MVT  and spectral lag) consistently place $G_1$ and $G_2$ in the short-like and long-like empirical regimes, respectively. Additional diagnostics based on derived spectral quantities, including the Amati and Norris relations, EHD, and the $\epsilon$ parameter, give the same qualitative distinction, while afterglow, supernova  and host-galaxy data provide contextual support for this interpretation. Taken together, GRB 160425A is neither a classical short-duration burst ($G_1$) with a later-time soft extended-emission component or X-ray flare ($G_2$), nor a typical long-duration burst ($G_2$) preceded by an early-time precursor ($G_1$). Instead, it is most naturally interpreted as an independent short-duration GRB event ($G_1$) followed by an independent long-duration GRB event ($G_2$) within the same event. The unique properties observed in GRB 160425A provide compelling evidence for a solid transition from a short-like (type I, merger origin) to a long-like (type II, collapsar origin) episode within a single system. As such, GRB 160425A challenges the traditional morphological classification scheme and our physical understanding of known GRB progenitor types, implying that SGRBs and LGRBs can occur almost simultaneously within the same astrophysical system, a phenomenon that has been neither observed nor predicted by current theoretical models.

\clearpage
\begin{table}
\refstepcounter{table}\label{tab:Global}
\setlength{\tabcolsep}{0.10em}
\renewcommand\arraystretch{1.2}
{\bf Table~1~~Properties of GRB 160425A}\\
\centering
\begin{tabular}{l|l}
\hline
{\bf Basic Observation}&{\bf value}\\
\hline
Duration (15-350 keV) [$T_{90}$]&304.6$\pm$15.0~(s)\\
1-s peak flux (15-150 keV) [$F_{\rm p}$]&(2.8$\pm$0.2)~(ph~cm$^{-2}$~s$^{-1}$)\\
Peak luminosity (15-150 keV) [$L_{\rm \gamma, p, iso}$]&(2.6$\pm$1.2)$\times$10$^{50}$~(erg~s$^{-1}$)\\
Total BAT fluence (15-150 keV) [$S^{\rm BAT}_{\gamma}$]&(2.1$^{+0.2}_{-0.2}$)$\times$10$^{-6}$~(erg~cm$^{-2}$)\\
Total bolometric fluence (1-10$^{4}$ keV) [$S_{\gamma}$]&(3.4$\pm$0.3)$\times$10$^{-6}$~(erg~cm$^{-2}$)\\
Total isotropic energy (1-10$^{4}$ keV) [$E_{\rm \gamma, iso}$]&(4.7$^{+0.7}_{-2.4}$)$\times$10$^{51}$~(erg)\\
Observed frame quiescent gap [$t_{\rm gap}$]&255.5$^{+9.3}_{-9.7}$~(s)\\ 
Rest frame quiescent gap [$t_{\rm gap}/(1+z)$]&164.3$^{+6.0}_{-6.2}$~(s)\\ 
\hline
{\bf Host Galaxy}\\
\hline
Redshift [$z$]&0.555\\
Luminosity distance [$D_{\rm L}$]&3308~(Mpc)\\
Offset [$R_{\rm off}$]&1.86~(kpc)\\
Half light radius [$R_{\rm 50}$]&2.5~(kpc)\\
Normalized offset [$r_{\rm off} \equiv R_{\rm off}/R_{\rm 50}$]&0.7\\
Cumulative light fraction [$F_{\rm light}$]&$\sim$0.6\\
Star formation rate [SFR]&0.58~($M_{\odot}$ yr$^{-1}$)\\
Optical transient location&18h~41m~18s \quad $-54^\circ 21^\prime 36^{\prime \prime}$\\
\hline
{\bf Associations}\\
\hline
Kilonova-association&Uncertain\\
SN-association&Uncertain\\
\hline
\end{tabular}
\end{table}

\clearpage
\begin{table}
\refstepcounter{table}\label{tab:IndBurst}
\setlength{\tabcolsep}{0.10em}
\renewcommand\arraystretch{1.2}
{\bf Table~2~~Individual burst properties of GRB 160425A}\\
\centering
\begin{tabular}{l|l|l|l|l}
\hline
Parameter & \multicolumn{2}{c|}{GRB 160425A-$G_1$} & \multicolumn{2}{c}{GRB 160425A-$G_2$} \\
\hline
&Value&Classification&Value&Classification\\
\hline
Time interval (15-350 keV) [$t_1$$\sim$$t_2$]&[-0.20$\sim$1.50]~(s)&&[257.0$\sim$308.2]~(s)&\\
BAT-determined position[RA(J2000.0), Dec(J2000.0)]&[$280.34^\circ$, $-54.36^\circ$]&&[$280.29^\circ$,$-54.35^\circ$]&\\
Duration (15-350 keV) [$T_{90}$]&1.7$^{+0.5}_{-0.4}$~(s)&Short&51.2$^{+8.9}_{-9.2}$~(s)&Long\\
Hardness ratio [$\frac{S(25-50 \ {\rm keV})}{S(15-25 \ {\rm keV})}$]&1.74$\pm$0.09&Short&0.40$\pm$0.04&Long\\
Minimum variability timescale (MVT)&18.5$\pm$3.1~(ms)&Short&550$\pm$32~(ms)&Long\\
$f$($f_{\rm eff}$)- parameter [$F_{\rm p}(F^{'}_{\rm p})/F_{\rm b}$]&2.13$\pm$0.18&Short&1.05$\pm$0.03&Long\\
Time lag [25-50 keV $\sim$ 15-25 keV]&(38$\pm$39)~(ms)&Short&(1152$\pm$264)~(ms)&Long\\
Spectral photon index [$\alpha$]&-1.75$\pm$0.08&&-1.17$\pm$0.02&\\
Spectral peak energy [$E_{\rm p}$]&$\gtrsim$500~(keV)&Short&57.6$\pm$2.7~(keV)&Long\\
Bolometric (1-10$^{4}$ keV) fluence [$S_{\gamma}$]&(8.6$\pm$1.1)$\times$10$^{-7}$~(erg~cm$^{-2}$)&Short&(2.5$\pm$0.3)$\times$10$^{-6}$~(erg~cm$^{-2}$)&Long\\
Peak luminosity (15-150 keV) [$L_{\rm \gamma, p, iso}$]&(2.0$\pm$0.2)$\times$10$^{50}$~(erg~s$^{-1}$)[64-ms] &Short&(5.8$\pm$0.4)$\times$10$^{49}$~(erg~s$^{-1}$)[1-s]&Long\\
Isotropic energy (1-10$^{4}$ keV) [$E_{\rm \gamma, iso}$]&(7.2$\pm$0.9)$\times$10$^{50}$(erg)&Short&(2.1$\pm$0.3)$\times$10$^{51}$~(erg)&Long\\
$\varepsilon$-parameter [$E_{\gamma,\rm iso,52}/E^{5/3}_{\rm p,z,2}$]&0.002$\pm$0.001&Short&0.25$\pm$0.03&Long\\
EHD-parameter [$E_{\rm p,i,2} E^{-0.4}_{\rm iso,51}T^{-0.5}_{90,i}$]&9.5$\pm$2.1&Short&(1.2$\pm$0.1)$\times$10$^{-2}$&Long\\
Amati relation [$E_{\rm p,z} \propto E^{0.5}_{\gamma,\rm iso}$]&Off the relation&Short&On the relation&Long\\
Norris relation [$L_{\gamma,\rm p, iso}\propto \tau^{-1.14}$]&Off the relation&Short&On the relation&Long\\
\hline
\end{tabular}
\end{table}

\clearpage
\setcounter{table}{2}
\begin{table}
\refstepcounter{table}\label{tab:GRBtype}
\scriptsize
\setlength{\tabcolsep}{0.50em}
\renewcommand\arraystretch{1.2}
{\bf Table~3~~A Phenomenological Classification of GRB Prompt Emission Light-Curve Morphologies.}\\
\centering
\begin{tabular}{lllllll}
\hline
\hline
Phenomenological class&Representative events&Detected rate&Quiescent gap?&Traditional model&Possible type&Reference\\
&(e.g.,)&&(Background-consistent)&(Obeyed/Violated)&(Type I vs. Type II)&(e.g.,)\\
\hline
{Non-quiescent GRBs}\\
(Single/overlapping emission episodes)\\
\hline
Short-duration merger-like event&GRB 130603B&Commonly&No&Obeyed&Traditional Type I&\cite{Tanvir2013}\\
Long-duration merger-like event&GRB 211211A&Rarely&No&Obeyed&Peculiar Type I&\cite{Rastinejad2022}\\
Continuous boundary/Hybrid event&GRB 180418A&Rarely&No&Obeyed&Type I or Type II&\cite{Becerra2019}\\
Short-duration collapsar-like event&GRB 200826A&Rarely&No&Obeyed&Peculiar Type II&\cite{ZhangBB2021}\\
Long-duration collapsar-like event&GRB 030329&Commonly&No&Obeyed&Traditional Type II&\cite{Hjorth2003}\\
\hline
{Quiescent GRBs}\\
(Two distinct emission episodes)\\
\hline
Precursor($G_1$)+LGRB($G_2$)&GRB 060124&Commonly&Yes&Obeyed&Traditional Type II&\cite{Romano2006}\\
Double-main/trigger LGRB($G_1$+$G_2$)&GRB 110709B&Commonly&Yes&Obeyed&Traditional Type II&\cite{ZhangBB2012}\\
SGRB($G_1$)+extended emission($G_2$)&GRB 050709&Commonly&Yes&Obeyed&Traditional Type I&\cite{Norris2006}\\
LGRB($G_1$)+late X-ray flare($G_2$)&GRB 050502B&Commonly&Yes&Obeyed&Traditional Type II&\cite{Falcone2006}\\
SGRB($G_1$)+late X-ray flare($G_2$)&GRB 050724&Commonly&Yes&Obeyed&Traditional Type I&\cite{Dai2006}\\
LGRB($G_1$)+afterglow onset($G_2$)&GRB 960530&Rarely&Yes&Obeyed&Traditional Type II&\cite{Hakkila2004}\\
{\bf SGRB($G_1$)+LGRB($G_2$)}&{\bf GRB 160425A}&{\bf The only one}&Yes&{\bf Violated}&{\bf Unclear}&{\bf This paper}\\
\hline
\end{tabular}
\vspace{0.5em}
\parbox{\linewidth}{\raggedright\scriptsize \textit{Note:} Bold type identifies GRB~160425A, the event analysed in this Article, and highlights the entries that distinguish its prompt-emission phenomenology from those of the other representative events listed in the table.}
\end{table}

\clearpage

\clearpage

\clearpage

\clearpage
\begin{center}
\textbf{Methods}
\end{center}

\noindent {\bf Uniqueness of GRB 160425A in the overall GRB population }\par
\noindent \label{sec:Uniqueness} GRB 160425A is unique in terms of the following three aspects. (1) It features two distinct emission episodes, an initial sub-burst, designated $G_1$, followed by a second, $G_2$. The two sub-bursts exhibit a quiescent interval measuring approximately 256 s in the observed frame and about 165 s in the rest frame, making GRB 160425A one of the longest quiescent gaps observed between GRB pulses. (2) The initial sub-burst, $G_1$, shows a sharp, single-peaked light curve and a short duration ($\sim$ 1.7 s), appearing like a traditional SGRB. Conversely, the subsequent sub-burst, $G_2$, exhibits a broad, single-peaked light curve and long duration ($\sim$ 51.2 s), aligning with the characteristic of a traditional LGRB. (3) During its prompt-emission phase, nearly all standard diagnostics for GRB classifications consistently categorize $G_1$ as a short-like (type I merger origin) event and $G_2$ as a long-like (type II collapsar origin) event. These diagnostics include key parameters such as duration, pulse shape, minimum variability, spectral lag, hardness ratio, EHD and $\varepsilon$ parameters, along with the Amati and Norris correlations. These compelling and consistent findings rule out alternative interpretations, such as the occasional occurrence of precursors or flares often observed during burst processes. These properties together make GRB 160425A a unique event, providing a solid identification for a transition from short-like (type I merger origin) event to long-like (type II collapsar origin) event within a single system.

\noindent {\bf Quiescent gap of GRB 160425A }\par
\noindent 
To assess whether the quiescent gap observed in GRB 160425A is unique, we compiled a comprehensive dataset of `quiescent' GRBs by combining data from two independent studies\cite{LanLin2018,LiLiande2022}. These studies systematically analysed quiescent times from  quiescent  GRB samples detected by \emph{Swift} and \emph{Fermi}. Our compiled dataset comprises 102 Fermi-detected GRBs\cite{LanLin2018} and 52 \emph{Swift}-detected GRBs\cite{LiLiande2022}. Figure \ref{fig:lcs_PE}b presents the distribution of quiescent times ($\Delta t_{\rm gap}$) within this dataset and GRB 160425A and other GRBs with long quiescent intervals highlighted within the distribution.

Remarkably, among all the bursts in the sample, only a handful of events exhibit a quiescent gap longer than or comparable to that of GRB 160425A. These events include GRB 041219A and GR 060124A, as well as three double-tracking GRBs: GRB 091024A, GRB 110709B  and GRB 220627A.

\noindent{\bf \emph{Swift} data retrieved }\par
\noindent 
We retrieve the \textit{Swift} data from the UK \textit{Swift} Science Data Centre (UKSSDC, \url{https://www.swift.ac.uk}). Data reduction was performed using HEASoft 6.30 and relevant calibration files (\url{http://heasarc.gsfc.nasa.gov/lheasoft/}). The Fermi Online Server (\url{https://apps.sciserver.org/dashboard/}) at the Fermi Science Support Center (FSSC, \url{https://fermi.gsfc.nasa.gov/ssc/}) was utilized for data processing. We conducted the extraction of photon events and subsequent temporal and spectral analysis following the methodologies outlined in refs. \citen{Evans2007,Evans2009}. Moreover, our analysis is double-checked by implementing the Multi-Mission Maximum Likelihood framework (3ML)\cite{Vianello2015} via the exported spectra nested in the 3ML package as a plugin.

\noindent{\bf Prompt-emission light-curve analysis }\par
\noindent  
Using the \emph{Swift}-BAT data obtained from the \emph{Swift} archive website (\url{https://www.swift.ac.uk/archive/ql.php}), we generated the \emph{Swift}-BAT light curve. HEASoft tools (version 6.30) was used following standard analysis threads (\url{https://www.swift.ac.uk/analysis/bat/}). We use \emph{batbinevt} to create the 1,024-ms light curves using a uniform-bin method across four standard \emph{Swift}-BAT energy bands: 15-25 keV, 25-50 keV, 50-100 keV  and 100-350 keV. The light curves for different energy bands are illustrated in Fig. \ref{fig:lcs_PE}, revealing a singular sharp-peaked structure ($G_1$) lasting approximately 1.7 s, succeeded by a broader peak ($G_2$) lasting $\sim$ 51.2 s at $T_{0}$+257.0~s, separated by a long quiescent period ($\sim$250~s). By computing the cumulative photon distribution across the entire BAT energy band (15-350 keV), we obtained   $T_{90}$ for both short-duration and long-duration bursts (Fig. \ref{fig:lcs_PE}), representing the duration between the epochs encompassing 5\% and 95\% of the total fluence. The 1$\sigma$ error region of the cumulative photon distribution was also determined by cumulatively summing the error of photon rate (counts per second per detector) for each time bin (indicated in pink in Fig. \ref{fig:lcs_PE}). From the sharp peak occurring between $T_{0}$-0.2 s and $T_{0}$+1.5~s of the short burst $G_1$, we computed $T_{90}\approx$ 1.7~s. Similarly, for the broad peak $G_2$ spanning $T_{0}$+257.0 s to $T_{0}$+308.2~s, we determined $T_{90}\approx$ 51.2~s. 

\noindent{\bf \emph{Swift}-BAT spectral data analysis }\par
\noindent  
With the BAT data downloaded from \emph{Swift} archive website, we then use \emph{batbinevt}, \emph{batupdatephakw}, \emph{batphasyserr}  and \emph{batdrmgen} to generate the spectral and relevant response files, including accumulated BAT event and detector plane histogram data into spectra (\emph{batbinevt}), updating BAT ray-tracing columns in spectral files (\emph{batupdatephakw}), applying the BAT spectral systematic error vector (\emph{batphasyserr}), and computing a BAT detector response matrix   for a known source position (\emph{batdrmgen}). Using the generated spectral and response files, we conducted time-integrated spectral analysis within the $T_{90}$ time intervals determined by the cumulative photon distribution for both $G_1$ and $G_2$. We used the typical spectral fitting models, such as the simple power-law (PL), cut-off power-law (CPL), Band function (Band)  and blackbody (BB), along with their hybrid versions to fit the spectra. Initially, spectral fitting was performed using XSPEC, followed by double-checking the fit results by implementing the 3ML via the exported spectra nested in the 3ML package as a plugin. For $G_1$ ($T_{0}$-0.20 s to $T_{0}$+1.50~s), the spectral fitting did not reveal a distinct peak within the BAT energy range, suggesting that the BB, CPL  or Band models may not adequately describe the data. Consequentially, the peak energy $E_{\rm p}$ must lie beyond the BAT energy range (15-150 keV) to avoid divergence of luminosity beyond this range. For $G_2$ ($T_{0}$+257.0 s to $T_{0}$+308.02~s), a spectral peak was observed in the fitting, with the Band model presenting the best fit. The resulting parameters included a low-energy photon index $\alpha=-1.17\pm0.10$, a high photon index $\beta=-2.44\pm0.08$  and the peak energy $E_{\rm p}=57.6\pm2.7$ keV. 
The photon number spectra of the PL and Band models can be described as
\begin{equation}
N(E) = A (\frac{E}{E_{\rm piv}})^{\Gamma_{\rm PL}},
\label{Eq:PL}
\end{equation}
and
\begin{eqnarray}
N(E)=A \left\{ \begin{array}{ll}
(\frac{E}{E_{\rm piv}})^{\alpha} \rm exp (-\frac{{\it E}}{{\it E_{0}}}), & E \le (\alpha-\beta)E_{0}\\
\lbrack\frac{(\alpha-\beta)E_{0}}{E_{\rm piv}}\rbrack^{(\alpha-\beta)} \rm exp(\beta-\alpha)(\frac{{\it E}}{{\it E_{\rm piv}}})^{\beta}, & E\ge (\alpha-\beta)E_{0}\\
\end{array} \right.
\label{eq:Band} 
\end{eqnarray}
respectively, where $A$ is the normalization factor at 100 keV in units of photons per cm$^2$ per keV per second, $E_{\rm piv}$ is the pivot energy fixed at 100 keV, $\Gamma_{\rm PL}$ is the power-law photon index, and $\alpha$ and $\beta$ are the low- and high-energy power-law photon spectral indices, respectively. $E_{\rm p}$ represents the peak energy, which is related to $E_{0}$ through the equation $E_{\rm p}=(2+\alpha)E_{0}$.

\noindent{\bf \emph{Swift}-XRT spectral data analysis }\par
\noindent  The \emph{Swift}-XRT data were acquired from the \emph{Swift} archive website, focusing on the time-sliced spectrum between $T_{0}$+257.0 s and $T_{0}$+308.2~s, corresponding to the duration of $G_2$. This time interval was chosen to ensure consistency with the temporal characteristics of the long burst, as no X-ray counterparts were observed during the epoch of $G_1$. The \emph{Swift}-XRT instrument provides high-resolution X-ray data of GRB afterglows. We first retrieve its raw data from UKSSDC, followed by the data reduction procedure using the HEASoft 6.30 and relevant calibration files. The position of the source is determined by fitting the X-ray point spread function, or taken from the Ultraviolet and Optical Telescope (UVOT) observation if it exists; approximately 75\% of positions are enhanced by the UVOT observation (\url{http://www.swift.ac.uk/xrt_positions/}). XSELECT is adopted to extract images, light curves  and spectra\cite{Evans2007,Evans2009}, and the related pile-up effect is corrected if necessary. We follow the suggestion by {\it the Swift} data reduction guide (\url{http://www.swift.ac.uk/analysis/})  that the threshold of applying pile-up correction is given as, for windows timing data, a count rate   above 100~counts per second, and for photon-counting data, a count rate exceeding 0.5~counts per second; the detailed method follows ref. \citen{Romano2006}. XSPEC is adopted for fitting the spectrum, which is binned as at least 1 count per bin by GRPPHA to ensure the validity of Cash statistics\cite{Cash1979}. In general, a single power-law model combined with the hydrogen absorption from the host galaxy and our Galaxy is preferred, but for rigour, we also check for the existence of an extra thermal component using the maximum likelihood ratio method to select one from two nested models by comparing the statistical significance. The uncertainty of the fitted parameters is estimated for a 90\% confidence level by the ERROR command and examined by the Markov chain Monte Carlo   iteration using the CHAIN command. 

\noindent Our refined spectral analysis revealed that the \emph{Swift}-XRT data during the X-ray flare phase (between 257.0 s and 308.2 s, corresponding to the epoch of $G_2$) could be accurately described by a simple power-law model, yielding a power-law index of $\Gamma=1.75\pm0.05$. Subsequently, the X-ray light curve was transformed into its corresponding cosmological rest frame, with appropriate $k$-correction applied\cite{Bloom2001}. Further details regarding the XRT automated analysis methods can be found in refs. \citen{Evans2007,Evans2009,Evans2014}.

\noindent{\bf Fermi-GBM signal search and data analysis (untriggered) }\par
\noindent  
We obtained the position history (POSHIST) data of the Fermi satellite from the Fermi-GBM Daily Data archive (\url{https://heasarc.gsfc.nasa.gov/W3Browse/fermi/fermigdays.html}) for 25 April   2016 (Extended Data Fig.~3). At the time when \emph{Swift}-BAT was triggered, 23:26:11 UT, the Fermi satellite and GRB 160425A were located on opposite sides of   Earth. In other words, the GRB was blocked by   Earth, and the Fermi satellite was unable to observe it (Extended Data Fig.~3a; the red pentagram represents GRB 160425A  and the blue region represents   Earth). We also plotted the light curve of all-sky observed by Fermi (Extended Data  Fig.~3b; there is no clear signal 5 min before and after the trigger time).

\noindent{\bf Hardness ratio }\par
\noindent 
It has been observed that SGRBs tend to exhibit harder spectra compared with typical LGRBs\cite{Dezalay1992}. To quantify this, we define a hardness ratio as the observed photon counts ratio between the 25-50 keV and 15-25 keV energy bands. For $G_1$, we compare the observed counts between $T_{0}$-0.20 s and $T_{0}$+1.50~s, while for $G_2$, the comparison is made between $T_{0}$+257.0 s and $T_{0}$+308.2~s. The individual bursts of GRB 160425A are indicated by a coloured star in Fig. \ref{fig:Classification}a, while the {\it Swift} GRB sample\cite{Horvath2010}, categorized into SGRB and LGRB populations\cite{Kouveliotou1993}, is represented by coloured points. As depicted in Fig. \ref{fig:Classification}a, $G_1$ is situated within the 1$\sigma$ region of the SGRB population whereas $G_2$ does fall within the 1$\sigma$ region of the LGRB population. 

\noindent{\bf Minimum variability timescale}\par
\noindent \label{sec:Timescale}
The MVT ($\Delta t_{\rm min}$) represents the shortest observable timescale over which a GRB exhibits significant flux variations. This metric provides insights into the minimum size of the emitting region and the underlying energy dissipation physics. For LGRBs and SGRBs, the median MVTs based on a Fermi burst sample are 134~ms and 18~ms, respectively\cite{Golkhou2015}. We determine the MVT for each individual burst of GRB 160425A by applying the technique utilizing Haar wavelets to the mask-weighted \emph{Swift}/BAT light curve with 1~ms binning, following the procedure in ref. \citen{Golkhou2014}. Our analysis reveals an MVT of $18.5 \pm 3.1$~ms for $G_1$ and $550\pm32$~ms for $G_2$ (Table \ref{tab:IndBurst} and Fig. \ref{fig:Classification}). This suggests that $G_1$ is consistent with the properties of SGRBs (type I), while $G_2$ aligns with the characteristics of LGRBs (type II).

\noindent{\bf Amplitude parameter }\par
\noindent 
The `amplitude' parameter $f$, defined by $f\equiv \frac{F_{\rm p}}{F_{\rm b}}$, has been proposed as a criterion to differentiate a short-duration burst serving as a `tip of the iceberg' of a long-duration burst or being a genuinely short-duration burst\cite{Lv2014}, where $F_{\rm p}$ and $F_{\rm b}$ are the peak flux and average background flux of the GRB light curve, respectively. According to statistical findings based on a large {\it Swift} GRB sample, a disguised SGRB would have $f<$1.5 (ref. \citen{Lv2014}). In GRB 160425A, we obtain $f=2.13\pm0.18$ for $G_1$ (refer to Table \ref{tab:IndBurst} and Fig. \ref{fig:lcs_PE}), a value that lies outside the range associated with disguised SGRBs. This observation strongly suggests that $G_1$ exhibits characteristics consistent with genuine SGRBs.

The tip-of-the-iceberg effect can disguise some type II LGRBs as apparently type I SGRBs. To distinguish intrinsically type I SGRBs, an effective amplitude parameter $f_{\rm eff}\equiv \frac{F^{'}_{\rm p}}{F_{\rm b}}= \xi f$ can be defined for type II LGRBs by quantifying this effect. Here, $F^{'}_{\rm p}$ is the peak flux of a pseudo-GRB with amplitude lowered by a factor $\xi$ from an original LGRB so its duration falls under 2 s. Typically, type II LGRBs have systematically smaller $f_{\rm eff}$ than the type I SGRBs. Following the procedure in ref. \citen{Lv2014}, we determine $f_{\rm eff}=1.05\pm0.03$ for $G_2$ of GRB 160425A, as listed in Table \ref{tab:IndBurst}. This small $f_{\rm eff}$ aligns with known values for LGRBs.

\noindent{\bf Spectral lag }\par
\noindent 
Spectral lag, as defined by ref. \citen{Norris2000}, refers to the time difference in the arrival of $\gamma$-ray photons of different energies emitted during a GRB. This phenomenon, observed in the majority of GRBs, serves as a distinguishing factor between the characteristics of long-duration and short-duration bursts. Typically, a time delay is observed between the light curves of two energy bands, with the higher-energy band peaking earlier than the lower-energy band. Observations show that type I SGRBs exhibit a negligible or even negative spectral lag \cite{Gehrels2006,Norris2006,2005Natur.437..855V,2014MNRAS.442.2342D,2010ApJ...725.1202L,2011ApJ...735...23N} whereas type II LGRBs are associated with a non-negligible spectral lag\cite{Yi2006, ZhangZhibin2006, Bernardini2015}. We use the cross-correlation function (CCF)\cite{Cheng1995, Band1997,Norris2000,Ukwatta2010}  analysis method to assess the time-averaged spectral lag and its uncertainty between the light curves of the softest band and progressively harder band for each individual burst of GRB 160425A (Fig. \ref{fig:Classification}c-e). The modified CCF derived by ref.\citen{Band1997}, without mean subtraction, is better suited for transient events such as GRBs. For two count rate series $x_{i}$ and $y_{i}$, where i=0,1,2, … ($N$-1), the modified CCF can be written as
\begin{equation}
{\rm CCF}(k \Delta t; x, y)=\frac{\sum_{i={\rm max}(1,1-k)}^{{\rm min}(N,N-k)} x_{i}y_{(i+k)}}{\sqrt{\sum_{i}x_{i}^{2}\sum_{i}y_{i}^{2}}},
\end{equation}
where $\Delta t$ represents the time bin duration, $k \Delta t$ ($k$= …, -1, 0, 1, …) denotes the time delay, and $x_{i}$ and $y_{i}$ are the count rates in energy bands $E_{1}$ and $E_{2}$. 

In our analysis, we consider four standard {\it Swift}-BAT energy bands: 15-25~keV (channel 1), 25-50~keV (channel 2), 50-100~keV (channel 3)  and 100-350~keV (channel 4). For $G_1$, using a 32-ms binned light curve, we calculate the spectral lag between $T_{0}$-0.2~s and $T_{0}$+1.5~s across these standard energy bands. For $G_2$, a 256-ms binned light curve is used, and the spectral-lag analysis between $T_{0}$+257.0 s and $T_{0}$+308.02~s is conducted. Despite obtaining non-zero spectral lags for $G_2$, these values are closer to those of LGRBs. However, the spectral lags from $G_1$ are notably close to zero (tiny spectral lags) and much smaller than those from $G_2$ (Fig. \ref{fig:Classification}d,e and  Table \ref{tab:IndBurst}). 

\noindent{\bf Norris correlation }\par
\noindent 
We measured the isotropic peak luminosity of each individual burst of GRB 160425A and over-plotted the $L_{\rm p,iso}-\tau$ diagram (Norris relation) using a GRB sample studied in ref.\citen{Norris2000}, and the  100-300 keV  to  25-50 keV  spectral lag is adopted. LGRBs typically align with the Norris relation, whereas SGRBs tend to deviate from it. The individual bursts of GRB 160425A are denoted by coloured stars in Extended Data Fig. 1a, while the GRB sample in ref.\citen{Norris2000} is represented by coloured points. We observe that $G_2$ follows the Norris relation, whereas $G_1$   deviates markedly from it, suggesting distinct origins between $G_1$ and $G_2$.

\noindent{\bf Estimate the $E_{\rm p}$ of $G_{1}$}\par
\noindent 
As {\it Swift}-BAT operates within a relatively narrow energy band (15-150 keV) in its detectors, the spectrum of $G_{1}$ ($T_{0}$-0.20 s to $T_{0}$+1.50~s) is best fitted by a simple power-law model within the BAT energy band, yielding a normalized coefficient $A$=14.6$^{+3.5}_{-3.5}$ and power-law index $\alpha$=-1.75$\pm$0.06. Consequently, the peak energy $E_{\rm p}$ of $G_{1}$ must lie beyond the BAT band ($E_{\rm p}>$150 keV). 

We developed an approach to infer a robust lower limit on $E_{\rm p}$ when it lies beyond the instrument's upper energy threshold\cite{Li2026a}. Our procedure to estimate $E_{\rm p}$, along with the corresponding bolometric fluence ($S_{\gamma}$) and $k$-corrected (1-10$^{4}$ keV) isotropic energy released in the $\gamma$-ray band ($E_{\gamma, \rm iso}$), involves the following steps. (1) To estimate a lower limit on $E_{\rm p}$, we fix $E_{\rm p}$ at a sequence of values extending beyond BAT's upper limit (for example, from 150 keV up to several megaelectronvolts; Extended Data Fig.~2a). For each fixed $E_{\rm p}$, we perform Markov chain Monte Carlo   sampling to explore the posterior distributions of the remaining free parameters ($A$, $\alpha$  and $\beta$). (2) For each fixed $E_{\rm p}$, we compute several statistical criteria to assess the goodness of fit while accounting for model complexity. We analyse how these criteria vary with the fixed $E_{\rm p}$. As $E_{\rm p}$ increases beyond the true peak energy, improvements in fit quality diminish, resulting in a `bending point' or plateau in the criteria curves (Extended Data Fig.~2b). This bending point serves as a robust lower limit for $E_{\rm p}$. To estimate the uncertainty in the $E_{\rm p}$ lower limit, we consider the spread of bending points across different criteria. The posterior distributions from Markov chain Monte Carlo sampling also provide credible intervals for $A$, $\alpha$  and $\beta$ at each fixed $E_{\rm p}$, allowing us to assess parameter uncertainties comprehensively. (3) In practice, we applied this   method to estimate the lower limit for $G_{1}$. As shown in Extended Data Fig.~2b, the bending points in the information criteria curves were observed at $E_{\rm p} \approx 500$ keV for Akaike information criterion (AIC), Bayesian information criterion (BIC), deviance information criterion (DIC) and $\chi^2_\nu$. Therefore, we used $E_{\rm p}=500$ keV as the peak energy of $G_{1}$ for the following analysis (Extended Data Fig.~2c).

We also utilize the full Fermi/GBM burst sample to determine a potential upper limit for $E_{\rm p}$ of $G_{1}$. This analysis incorporates a complete GBM burst sample (before May 2024) comprising 3,762 bursts. We establish a separation line for $E_{\rm p}$ lying on the $3\sigma$ region of the $E_{\rm p}$ distribution for the full Fermi/GBM burst catalogue available at \url{https://heasarc.gsfc.nasa.gov/W3Browse/fermi/fermigbrst.html}. To accomplish this, we fit the $E_{\rm p}$ distribution for the full Fermi/GBM burst catalogue with log-normal distributions, $N(\mu,\sigma^{2})$, where $\mu$ and $\sigma$ represent the mean and standard deviation. The fits yield $N_{\rm GRBs}(2.24,0.44^{2})$ for the full GBM GRB sample. On the basis of these findings, we calculate the probability (denoted as $P_{\rm GRB}$) of $G_{1}$ belonging to the GRB population. At a confidence level   of 3$\sigma$, we obtain $P_{\rm GRB}=6.0\times 10^{-4}$ and $E_{\rm p}=3{,}648$~keV. This corresponds to a confidence level of $3\sigma$ and $P_{\rm GRB}=6.0 \times 10^{-4}$ for rejecting the GRB population (refer to Extended Data Fig.~2d). In Extended Data Fig.~2e, we present these three distinct $E_{\rm p}$ values ($E_{\rm p}=150$ keV, $500$ keV and $3{,}648$ keV) along with their corresponding spectral shapes, and illustrate their impact on the calculation of $S_{\gamma}$ (note that $S_{\gamma}$ is sensitive to the choice of $E_{\rm p}$), as shown on the $E_{\rm p,z}$-$E_{\gamma,\rm iso}$ diagram.
 
\noindent{\bf Amati correlation }\par
\noindent 
The correlation between the rest-frame peak energy $E_{\rm p,z}$ and the isotropic-bolometric-equivalent emission energy $E_{\gamma,\rm iso}=4\pi d_L^2 S_{\mathrm{gamma}} k_c/(1+z)$ of GRBs was initially discovered in ref. \citen{Amati2002}. Here  $d_{L}$ represents the luminosity distance, $S_{\gamma}$ denotes the energy fluence (in erg cm$^{-2}$), $k_c$ stands for the $k$-correction factor (adjusting the observed energy range to 1-10$^{4}$ keV), and $z$ signifies the redshift. It has been used to distinguish between the properties of long-duration and short-duration bursts, as SGRBs and LGRBs do not conform to the same Amati relation on the $E_{\rm p,z}$-$E_{\gamma,\rm iso}$ diagram\cite{Amati2002}. By utilizing GRB samples with known redshift\cite{Amati2002,Zhang2009b,Qin2013}, we over-plotted the $E_{\rm p,z}$-$E_{\gamma,\rm iso}$ diagram (Extended Data Fig.~1b). The luminosity distance was calculated using the standard $\Lambda$-CDM cosmological parameters set with $H_0=67.4$ km s$^{-1}$ Mpc$^{-1}$, $\Omega_{M}=0.315$  and $\Omega_{\Lambda}=0.685$ throughout the paper\cite{PlanckCollaboration2018}.

\noindent{\bf EH and EHD parameters }\par
\noindent 
Two parameters, energy-hardness EH=$E_{\rm p,i,2} E^{-0.4}_{\rm iso,51}$ and EHD=$E_{\rm p,i,2} E^{-0.4}_{\rm iso,51}T^{-0.5}_{90,i}$, were introduced for the classification of GRBs into type I and type II categories \cite{Minaev2020}. This classification method is particularly useful for GRBs with unknown redshifts. Here, $E_{\rm p,i,2}$ represents the rest-frame peak energy $E_{\rm p,i}$ in units of 100 keV, $E_{\rm iso,51}$ denotes the isotropic equivalent energy $E_{\gamma, \rm iso}$ in units of $10^{51}$ erg, and $T_{90,i}$ stands for the rest-frame duration in seconds. The most reliable parameter for blind type I and type II classification, as suggested in ref. \citen{Minaev2020}, is the EHD distribution. Type I GRBs tend to cluster in the high-EHD region, while type II GRBs are typically found in the low-EHD region, with a separation point occurring at EHD=2.6. We calculated EHD for each of the two sub-bursts of GRB 160425A and found that $G_1$ has EHD=9.5, whereas $G_2$ has EHD=0.01. Subsequently, we compared the two bursts on the EHD-$T_{90,i}$ diagram (Extended Data Fig.~1c). As a result, $G_1$ falls within the high-EHD region, while $G_2$ falls within the low-EHD region.

\noindent{\bf $\varepsilon$ parameter }\par
\noindent 
By introducing the parameter $\varepsilon=E_{\gamma,\rm iso,52}/E^{5/3}_{\rm p,z,2}$, a phenomenological classification method was introduced to distinguish the physical origins of GRBs, distinguishing between type I (compact-star origin) and type II (massive-star origin) GRBs. The two-class GRBs tend to cluster nicely in low-$\varepsilon$ ($\varepsilon<$0.03) and high-$\varepsilon$ ($\varepsilon>$0.03) regions, respectively, with a separation line at $\varepsilon\approx$ 0.03 using a GRB sample with known redshift\cite{Lv2010}. We calculated $\varepsilon$ for each individual burst of GRB 160425A and compared them on the $\log\varepsilon$--$\log T_{90,z}$ plane. It was observed that $G_1$ falls within the low-$\varepsilon$ region, while $G_2$ falls within the high-$\varepsilon$ region (Extended Data Fig.~1d). These results suggest that $G_1$ is probably of a type I origin, whereas $G_2$ should be of a type II origin.

\noindent {\bf Robustness of classification }\par
\noindent \label{sec:jointclass}
We quantify the robustness of the opposite classifications of $G_1$ and $G_2$ using the same SGRB and LGRB comparison samples plotted in the classification figures. For each diagnostic, the class likelihoods are evaluated with kernel density estimates (KDE)\cite{Rosenblatt1956,Parzen1962} of the observed samples, using Gaussian kernels with Scott's bandwidth rule\cite{Scott1992} as implemented in \texttt{SciPy}\cite{Virtanen2020}. The estimator centres a kernel on each comparison burst, so that the local density is set by the observed distribution of bursts and the empirical overlap of the two populations is preserved. Under equal class priors the support for the short-like class at a measured position $x$ is $R(S|x)=\hat{f}_S(x)/[\hat{f}_S(x)+\hat{f}_L(x)]$, where $\hat{f}_S$ and $\hat{f}_L$ are the SGRB and LGRB densities. 

Two limits were imposed. A comparison sample of $N$ bursts cannot constrain a probability below the level corresponding to a single object, so no diagnostic is allowed to exceed $N/(N+1)$. The discriminating power of each classifier is verified by leave-one-out cross-validation\cite{Stone1974}, in which each comparison burst is classified from the remaining bursts of both samples; 94-100\% are recovered in their published class. Diagnostics sharing an input are not treated as independent: the Amati, EHD  and $\varepsilon$ diagnostics all derive from $E_{\rm p}$ and $E_{\gamma,\rm iso}$ and are combined into a single energy group, the three spectral-lag band pairs into a single lag group, and duration is not counted separately because it is one axis of the hardness-duration plane; within a group, the log-odds are averaged rather than multiplied. The amplitude parameter is excluded from the combination because no uniform comparison sample is available for an empirical overlap estimate.

The resulting probabilities ($G_1$ short, $G_2$ long) are 0.91248 and 0.99507 for hardness-duration, 0.99210 and 0.98925 for MVT-duration, 0.99061 and 0.97429 for the energy group, 0.53690 and 0.68389 for spectral lag, and 0.94118 and 0.87500 for the lag-luminosity relation. Combining the five groups gives $P(G_1={\rm short})=1-3.9\times10^{-7}$ ($4.9\sigma$) and $P(G_2={\rm long})=1-9.3\times10^{-8}$ ($5.2\sigma$). No individual diagnostic exceeds 0.996, and the combined values follow from the concordance of 5 independent evidence groups, each bounded by its own comparison sample.

\noindent{\bf \emph{Swift}-XRT afterglow light curves }\par
\noindent \label{sec:AGLC}
SGRBs are known to exhibit lower luminosity compared with typical LGRBs in their afterglow light curve observed by \emph{Swift}-XRT data\cite{Zhang2018}. To differentiate the afterglow properties of GRB 160425A between SGRB and LGRB populations, a sample of 248 GRBs (217 LGRBs and 31 SGRBs) observed by \emph{Swift}-XRT with known redshift from 2004 to 2015 was compiled\cite{Li2015,Li2018b,Ruffini2018}. Using this sample, we over-plotted the luminosity as a function of rest-frame time on the $L_{\rm AG,iso}-t_{\rm z}$ diagram (Extended Data Fig.~4a), where $L_{\rm AG,iso}$ is the luminosity of the X-ray afterglow and $t_{\rm z}=t/(1+z)$ is the rest-frame time. In the diagram, GRB 160425A is indicated by yellow data points, while the GRB sample is depicted by black and grey lines for SGRB (type I) and LGRB (type II)   populations, respectively. Notably, GRB 160425A appears at the bottom-edge region of the long-burst distribution, but at the top-edge region of the short-burst distribution. These results suggest that we cannot definitively classify GRB 160425A as either a type I or type II $\gamma$-ray burst based solely on its afterglow emission.
 
\noindent {\bf X-ray afterglow observations }\par
\noindent  
The \emph{Swift}-XRT  began to observe the BAT field from 208~s to 1,171 ~ks after the BAT trigger. It identified a fading, uncatalogued X-ray source with an enhanced position located at right ascension (RA) =18h41 min 18.56s and declination (Dec) =-54$^{\circ}$21$^{\prime}$36.2$^{\prime \prime}$ (J2000) with an uncertainty of 2.0$^{\prime \prime}$. The initial data were obtained in 439 s in windowed-timing mode with the remainder in photon-counting mode. The X-ray light curve of GRB 160425A exhibits two distinct phases: an early-time X-ray flare and a late-time normal decay. We fit the X-ray flare (from 208 s and $\sim$ 600 s) with a smooth broken power-law model. The best-fit parameters give $\alpha_1=-43.90\pm3.91$, $\alpha_2=6.06\pm0.21$  and $t_{\rm p}=270\pm45$ s. The following normal decay (between $\sim$ 600~s and 1,171 ~ks)~  can be modelled with a power-law decay with a power-law index $\alpha_{\rm X}=1.03\pm0.06$ (Extended Data Fig.~4b). The results suggest that the early X-ray flare is consistent with the activation of the central engine, whereas the subsequent normal decay phase aligns with the standard external shock afterglow model. These light-curve fits are performed using the {\it lmfitt} \cite{Newville2016}, using a nonlinear least-squares method with the Levenberg-Marquardt minimization algorithm to fit the data. The convention $F_{\nu}\propto t^{-\alpha_{\rm X}} \nu^{-\beta_{\rm X}}$ is adopted, where $F_\nu$ is the flux density at frequency $\nu$, $\alpha_{\rm X}$ represents the temporal index of the light curve  and $\beta_{\rm X}$ is the photon index of the X-ray spectrum, with $\beta_{\rm X}=\Gamma_{\rm X}-1$.

\noindent {\bf Ultraviolet, optical and near-infrared afterglow observations }\par
\noindent  
At the coordinates provided by \emph{Swift}-XRT, we gathered ultraviolet, optical  and near-infrared observations (Extended Data Table~1). Optical data were   obtained from either General Coordinates Network (GCN) circulars (\url{https://gcn.gsfc.nasa.gov/other/160425A.gcn3}) by multiple facilities (European Southern Observatory (ESO) 2.2m MPG and CTIO 1.3m ANDICAM) or by the \emph{Swift}  UVOT\cite{Roming2005}  from the \emph{Swift} archive website (\url{https://www.swift.ac.uk/archive/ql.php})(Extended Data Table~1), together with our own observations performed by the Gamma-Ray Burst Optical/Near-Infrared Detector (GROND), attached to the MPG/ESO 2.2 m telescope at La Silla Observatory (Extended Data Table~2). To account for host galaxy and Galactic extinctions, as well as spectral $k$-corrections, corrections were applied to the observed data. Galactic extinction correction utilized the reddening map presented in ref. \citen{Schlafly2011} for optical and near-infrared magnitudes. The $k$-correction was applied to the observed spectra, defined as $m_{\rm T}=m_{\rm O}-k$, following refs.\citen{Oke1968, Peterson1997}, with $k=2.5(\beta_o-1)\log(1+z)$, where $m_{\rm O}$ and $m_{\rm T}$ are the observed and true magnitudes, respectively, $\beta_o$ is the spectral index measured from the optical afterglow (a typical value $\beta_o$=0.75 is applied), and $z$ is the redshift. Using a typical $A_{\rm v}$ value (${\rm log_{10}} A^{\rm host}_{\rm v}=-0.82\pm 0.41$ ; see Fig. 1 in ref. \citen{Li2018a}) and assuming $R_{\rm v}$=3.1, the host extinction correction in any optical band, $A_{\lambda}$, was calculated following the extinction curve defined in ref.\citen{Pei1992}, where $R_{\rm V}$ $\equiv$ $A_{\rm V}$/$E_{\rm B-V}$ is the total-to-selective extinction ratio, and the colour excess $E_{\rm B-V}$ $\equiv$ $A_{\rm B}$-$A_{\rm V}$ represents the difference between the extinction in B and V filters.

\noindent{\bf Unable to determine supernova }\par
\noindent \label{sec:SN} 
\noindent The observation of supernovae association is one of the most confident pieces of evidence for the progenitor of a GRB. It is generally accepted that type I bursts are formed by the merger of binary compact stars without the presence of a supernova component. Type II  bursts are generated by the collapse of massive stars, with nickel in their outwards expanding shells releasing energetic photons through nuclear radioactive decay\cite{Zhang2018}. These energetic photons are converted to optical emission escaping the opaque shells through diffusion for about 2 weeks\cite{2017AdAst2017E...5C}. Hence, the determination of the supernova component that appears as a bump on the optical light curve gives a possible judgement on the type II  burst.

\noindent GRB 160425A has extremely limited optical data observations, with the majority of bands having only a single time bin. The \emph{Swift} observations of the U band that we heavily rely on have values in two time bins and some time bins of upper limits. The U band is the best optical observational band available. It is essential to apply extinction corrections and subtract the host galaxy's background luminosity. The host galaxy of GRB 160426A has been localized, with observations from telescopes such as CTIO, MASTER and VLT\cite{Tanvir2016}. VLT/FOcal Reducer and low dispersion Spectrograph (FORS) reveals three distinct diffuse knots, suggesting   merging galaxies (Fig. \ref{fig:HG_image}). Despite these observations, the U-band magnitude of the host galaxy remains unavailable. Late-time CTIO observations report an I-band magnitude of $20.1$, with no detectable brightness variation (uncertainty $<$ 0.05 mag) between 7 days and 26 days post-burst. A supernova similar to SN 1998bw would have brightened the I-band by 0.1 magnitudes, which was not observed. These findings lead to the conclusion that if this is an LGRB, any associated supernova must be dimmer than SN 1998bw. For typical star-formation galaxies, the magnitude of $U-I$ is around 1.0 to 2.0 (ref. \citen{Calzetti1997AJ}); hence,   we assume from the galaxy's U-band brightness ($\sim 21.6\pm0.5$ mag) that the corresponding flux density is approximately $0.008$ mJy, which is close to the observed U-band flux density. This means after subtracting the host galaxy's contribution, the U-band flux would be very faint, as shown in Extended Data Fig.~6. Owing to the uncertainty in the brightness of the host galaxy, the error in the U-band data points has increased, particularly for the last data point, where the minimum value has already fallen below the flux of synchrotron radiation. The final \emph{Swift}/UVOT $U$-band data point at $t\approx10^6$ s is faint and, within the large uncertainty of the host estimate, is consistent with host-galaxy light. Because no direct late-time $U$-band host image or multi-band host spectral energy distribution is available, the host subtraction remains uncertain. The existing $U$-band data therefore cannot be used to isolate a supernova component.

Although it is uncertain, we still adopt SN 1998bw as a template light curve\cite{Clocchiatti2011}, trying to limit the supernova component by adjusting the redshift and scaling the luminosity and time\cite{2004ApJ...609..952Z}. We obtain the quasi-bolometric unreddened rest-frame light curves of SN 1998bw from ref. \citen{Clocchiatti2011}. To generate a particular supernova light curve, we apply a scaling factor, $k$, and a stretch factor, $s$, to the template supernova, SN 1998bw. At redshift $z$, the observed flux density of the specific supernova can be given by:
\begin{eqnarray}
F^{\rm SN}(t_{\rm obs},\nu_{\rm obs})=\frac{(1+z)k}{4\pi d^{2}_{\rm L}} L^{\rm SN}_{\nu}
\left(\frac{t_{\rm obs}}{(1+z)s},(1+z)\nu_{\rm obs}\right)\label{eq:supernova}
\end{eqnarray}
where $d_{\rm L}$ is luminosity distance and $L^{\rm SN}_{\nu}$ is specific luminosity at frequency $\nu$ of template supernova. To obtain plausible results from the limited data, we use numerical simulations of the forward shockwave model to derive the synchrotron component of the afterglow \cite{2020ApJ...896..166R}. In Extended Data Fig.~ 6, we present the multiwavelength observations of GRB 160425A alongside the fitted Gaussian jet model. Parameters include isotropic equivalent energy of the jet set to $E_0 = 0.9 \times 10^{53} \, \text{erg}$. The circumburst medium density is $n_0 = 0.001 \, \text{cm}^{-3}$  and the electron-energy distribution index is $p = 2.15$. The microphysical parameters are $\epsilon_e = 0.28$ and magnetic energy fraction ($\epsilon_B$) $=0.001$. To estimate the maximum luminosity of a hypothetical supernova under the constraints of the data, we found $s = 0.9$ and $k = 1$. This indicates that, if a supernova is present, its upper-limit brightness is at least one magnitude fainter than that of SN 1998bw. This result is consistent with the aforementioned CTIO observations.

At $z=0.555$, a GRB-associated supernova with a luminosity comparable to SN 1998bw is in principle observable with sufficiently deep and well-sampled optical data. For example, GRB 101219B at $z=0.5519$ was associated with SN 2010ma, whose luminosity was about 1.4 times that of SN 1998bw. In GRB 160425A, however, the available late-time optical data are sparse. The \emph{Swift}/UVOT $U$ band probes the rest-frame near-ultraviolet, where broad-lined type Ic supernovae are faint and template extrapolations are uncertain, while the available $I$-band data do not provide a clean host-subtracted supernova light curve. We therefore cannot confirm a supernova component.

\noindent We conclude that the data are insufficient to confirm the existence of a supernova, suggesting there is either no supernova component or a very weak one.

\noindent{\bf The BAT-determined position of $G_1$ and $G_2$ }\par
\noindent \label{sec:BATposition}
We adopt the HEASoft tools \texttt{batbinevt}, \texttt{batfftimage}  and \texttt{batcelldetect}, which analyse the event file, generate images  and derive source positions for each of the spikes. For the first spike $G_1$, we determined the centroid to be at RA = $280.34\degree$ and Dec = $-54.36\degree$, and the positional error is $0.02\degree$. For the second spike $G_2$, we determined the centroid to be at RA = $280.29\degree$ and Dec = $-54.35\degree$, and the error is $0.02\degree$. To check whether the two spikes share the same origin, we then compute the angular separation between the centroids and the combined positional error. The angular separation is $0.03\degree$ or about $1.69$ arcmin. The combined positional error is $0.02\degree$ or $1.42$ arcmin. Our results show that the positions of the two spikes are consistent within the uncertainties. 

The interval between $G_1$ and $G_2$ is not devoid of coverage. MASTER-SAAO reported a marginal optical transient of $\sim$18.2$\pm$0.5~mag in its first 30-s exposure at $T_0+168$-$198$~s (refs. \citen{2016GCN.19344....1L,2016GCN.19351....1B}), within $0.4''$ of the enhanced XRT position\cite{2016GCN.19347....1B} and hence of the $z=0.555$ host. The XRT detections from $T_0+208.8$~s onwards likewise originate from the counterpart position, but rise continuously into $G_2$ and are most naturally its early development. Neither observation isolates a component-resolved afterglow of $G_1$, so the redshift-independent prompt-emission diagnostics of $G_1$ are unaffected by the host association.

We conclude that the two emission episodes in GRB 160425A probably come from the same astrophysical source. 

\noindent{\bf Host-galaxy observations and properties}\par
\noindent \label{sec:HG}
It is imperative to consider the properties of the host galaxy when endeavouring to classify GRBs from a physical standpoint. From analysis of emission lines such as H$\alpha$, H$\beta$, [O III], [O II]  and [S III], the spectrum indicates two different systems at $z$=0.555. The observed velocity separation amounts to 180 km s$^{-1}$, while the spatial separation aligns with the 3 knots visible in the imaging (corresponding to 8 kpc at this distance), indicating an interacting pair of galaxies\cite{Tanvir2016}.

{\it Spectrum.} The host galaxy of GRB 160425A was observed with VLT/X-shooter on 26 April 2016, as part of the XSGRB project\cite{2019A&A...623A..92S}. The reduced spectrum is available at \url{https://sid.erda.dk/share\_redirect/DBuNORk1lI/XSGRB.zip} in the standard ESO format. We fit the emission lines with   GLEAM\cite{2021AJ....161..158S}. We find that the H$\alpha$ flux is $(0.8 \pm 0.2) \times 10^{-16}$ ergs s$^{-1}$ cm$^{-2}$. After correcting the Milky Way extinction\cite{Schlafly2011}, the luminosity is $(1.1 \pm 0.3) \times 10^{41}$ erg s$^{-1}$. The X-shooter spectrum is a combination of the afterglow and the host galaxy. The continuum is dominated by the afterglow, but the narrow nebular emission lines at the host redshift are attributed to the host galaxy. We therefore fitted and subtracted the local continuum around each line before measuring the line fluxes. Because the host is an interacting system and the spectrum may not capture the total integrated host light, the derived star formation rate (SFR) should be regarded as an approximate value, or as a lower limit to the total host-galaxy SFR. With the H$\alpha$ luminosity, the  SFR  of the host galaxy is estimated to be SFR $= L_{\rm H\alpha}/1.86\times 10^{41}$ $M_{\odot}$ yr$^{-1}$ = 0.58 $M_{\odot}$ yr$^{-1}$ (ref.~\citen{2012ARA&A..50..531K}). We estimate the metallicity with $\rm 12 + log (O/H) = 8.73 - 0.32 \times O3N2$, where $\mathrm{O3N2}=\log\{([\mathrm{O\,III}]\lambda5007/\mathrm{H}\beta)/([\mathrm{N\,II}]\lambda6583/\mathrm{H}\alpha)\}$ (ref.~\citen{2004MNRAS.348L..59P}). We determine the metallicity to be 8.53, a metal poorer than solar metallicity.

{\it Image.}  \noindent VLT/FORS2 observations of the field of GRB 160425A were obtained on multiple occasions between 13 May and 3 September 2016. To minimize contamination from the afterglow, we used only the images obtained on 3 September 2016 for the host-galaxy analysis. We used {\it ESO Reflex}\cite{2013A&A...559A..96F} (\url{https://www.eso.org/sci/software/esoreflex/}) for bias subtraction and sky-flat-field correction. Sources were detected with {\it SExtractor}\cite{1996A&AS..117..393B} (\url{https://www.astromatic.net/software/sextractor/}), after which the images were astrometrically registered with {\it SCAMP}\cite{2006ASPC..351..112B} (\url{https://www.astromatic.net/software/scamp/}) and resampled with {\it SWarp}\cite{2002ASPC..281..228B} (\url{https://www.astromatic.net/software/swarp/}) to a pixel scale of $0.125^{\prime\prime}$. The image of the host galaxy is presented as panels of Fig. \ref{fig:HG_image}. We registered the \emph{Swift}/UVOT white band image observed on 25 April   2016 to the Gaia Data Release 2 catalogue and detected the afterglow to be 18:41:18.5562, -54:21:36.416. After registering the VLT/FORS image to the \emph{Swift}/UVOT image, the centre of the host galaxy was detected as 18:41:18.5247, -54:21:36.334. Thus, the offset from the afterglow to the centre of the host is $0.29^{\prime\prime}$, corresponding to $R_{\rm off}=1.86$ kpc. 

We fitted the host galaxy with GALFIT (\url{https://users.obs.carnegiescience.edu/peng/work/galfit/galfit.html}) and determined $R_{50}$ to be $0.39^{\prime\prime}$, corresponding to 2.5 kpc at $z=0.555$. The normalized offset $r_{\rm off}=R_{\rm off}/R_{\rm 50}=0.7$, where $R_{\rm 50}$ is the half-light radius. The cumulative light fraction $F_{\rm light}\approx0.6$.

To assess whether the host environment of GRB 160425A is more consistent with the SGRB or LGRB populations, we compare its host-galaxy properties with published datasets for both classes\cite{Li2016,Fong2022}. We consider redshift,  SFR, physical offset ($R_{\rm off}$), half-light radius ($R_{\rm 50}$), normalized offset ($r_{\rm off}$)  and cumulative light fraction ($F_{\rm light}$), using only secure measurements and excluding upper and lower limits. For each physical parameter, we construct the corresponding distribution using a large catalogue of GRB host-galaxy data compiled from the literature~\cite{Li2016,Fong2022}. When the parameter distributions are sufficiently constrained, we fit the short- and long-duration GRB samples separately with a log-normal function, $\mathcal{N}(\mu,\sigma^2)$, in logarithmic space. We then evaluate the probability of the observed value of GRB~160425A under each fitted distribution and convert this probability into an equivalent Gaussian significance. For $F_{\rm light}$, where a single parametric fit is poorly constrained, we instead use  KDE. The full results are given in Extended Data Table~3   and Extended Data Fig.~7.

Taking the physical offset $R_{\rm off}$ as an example, we demonstrate our procedure for evaluating the host environment of GRB 160425A. We first combine the LGRB sample from ref.~\citen{Li2016} with the larger SGRB sample from ref.~\citen{Fong2022}, yielding 132 LGRB and 83 SGRB hosts with well-measured offsets. Gaussian fits to the log-transformed offset distributions give $\mathcal{N}_{\rm SGRB}(0.91,0.68^2)$ and $\mathcal{N}_{\rm LGRB}(0.29,0.62^2)$. GRB 160425A has a measured offset of $R_{\rm off}=1.86$~kpc, giving $p_{\rm SGRB}=0.17$ and $p_{\rm LGRB}=0.49$, corresponding to rejection levels of $0.94\sigma$ and $0.03\sigma$, respectively. The offset therefore favours an LGRB host, although it still lies within the SGRB range. We apply the same procedure to the remaining parameters: redshift, SFR, $R_{\rm off}$  and $r_{\rm off}$. For $F_{\rm light}$, neither population is well described by a single Gaussian, so we instead rely on a KDE. At $F_{\rm light}\approx0.6$, the KDE-derived probability densities are 0.57 for SGRBs and 0.86 for LGRBs, again favouring an LGRB environment.

Overall, the host-galaxy diagnostics give no clean classification. The redshift and SFR show a slight preference towards SGRB hosts, while $R_{\rm off}$, $r_{\rm off}$  and $F_{\rm light}$ are more consistent with LGRB hosts, and $R_{\rm 50}$ remains uninformative.

\noindent{\bf GRB 160425A in synchrotron afterglow emissions of \emph{Swift} GRBs}\par
\noindent 
We compare the afterglow synchrotron parameters of GRB 160425A with other GRBs.

We constructed a sample of \emph{Swift}-XRT  GRBs  with known redshift. Light curves were retrieved from the UKSSDC and included both windowed-timing and photon-counting modes. We excluded incomplete light curves that had no observations before 1,000 s, and also excluded those with fewer than 10 data points in the default binning on the UKSSDC website. In total, 147 GRBs were selected. For a homogeneous population comparison, all bursts in this sample were modelled with the \emph{Swift}-XRT afterglow data only. This keeps the data selection and fitting procedure identical across the full sample, as optical and radio coverage is not available with comparable cadence and calibration for most events. Early-time flares and late-time bumps were excluded to isolate the forward-shock afterglow emission. To improve the homogeneity of the dataset and reduce statistical imbalance at late times, we applied a logarithmic rebinning in observer time. The observed 0.3-10 keV fluxes were converted into monochromatic flux density at 1 keV, assuming a power-law photon spectrum.

The afterglow emission was modelled with a semi-analytic synchrotron forward-shock code, \texttt{jetsimpy}\cite{2024ApJS..273...17W}, which computes broadband light curves for both constant-density interstellar medium (ISM) and stellar wind density profiles. The physical parameters of interest were the isotropic equivalent kinetic energy of the blast wave, $E_{\mathrm{k,iso}}$; the initial bulk Lorentz factor of the outflow, $\Gamma_0$; the normalization of the circumburst number-density profile, $n_{\mathrm{env}}$ (denoted $n_0$ for the constant-density ISM case); the fractions of the post-shock internal energy carried by accelerated electrons and magnetic fields, $\epsilon_e$ and $\epsilon_B$, respectively; and the power-law index of the non-thermal electron-energy distribution, $p$. Log-uniform priors were adopted for energy and microphysical fractions, and uniform priors for $p$. Posterior distributions were obtained using the affine-invariant Markov chain Monte Carlo sampler \texttt{emcee}, with convergence monitored via autocorrelation analysis. Parameter estimates are reported as posterior medians with 68\% credible intervals bounded by the 16th and 84th percentiles. The joint posterior sampling accounts for covariances among the fitted parameters within the adopted model. Representative fits were also checked with \texttt{VegasAfterglow}\cite{2026JHEAp..5000490W} and \texttt{afterglowpy}\cite{2020ApJ...896..166R}; these tests gave consistent parameter ranges and the same placement of GRB~160425A relative to the comparison sample. To assess the circumburst environment, we compared ISM and wind fits using the  DIC, with $\Delta\mathrm{DIC}\gtrsim5$ considered substantial evidence.

The derived distributions reveal commonalities and systematic differences between SGRBs and LGRBs. The electron-energy fraction clusters tightly around $\epsilon_{e} \approx 0.2$-$0.3$, with LGRBs showing values modestly higher than short-duration bursts. The electron power-law index is stable across the sample, with $p \approx 2.2$-$2.3$, consistent with relativistic shock acceleration theory. In contrast, the magnetic energy fraction spans several orders of magnitude, $\epsilon_{B}\approx 10^{-4}$-$10^{-1}$, with medians of a few $\times 10^{-3}$. This wide dispersion suggests variable efficiency in magnetic-field amplification. This distribution is conditional on the adopted data selection and forward-shock modelling framework, whose parameters may be degenerate. For example, an estimate based on the radio peak times and fluxes of 36 GRB afterglows found a narrower distribution centred at $\epsilon_e\approx0.13$-$0.15$ (ref.~\citen{Beniamini2017}). By contrast, Bayesian broadband modelling of 22 LGRBs and 4 SGRBs found $\epsilon_e$ values spanning approximately $0.1$-$1$, with mean values of $0.34$ and $0.28$ for homogeneous and wind-like environments, respectively\cite{Aksulu2022}.

Energetics differ strongly between populations: LGRBs typically exhibit $E_{\mathrm{k,iso}}$ $\approx10^{52}$-$10^{54}\,\mathrm{erg}$, while SGRBs cluster near $5\times10^{51}\,\mathrm{erg}$, with no event exceeding $2\times10^{52}\,\mathrm{erg}$. The latter is comparable to the maximal spin-down energy of a millisecond magnetar, indicating a natural ceiling on the afterglow energy budget. The initial Lorentz factors also separate the two populations: SGRBs peak at $\Gamma_{0}\approx 300$-$400$, exceeding the typical values of LGRBs ($\Gamma_{0}\approx 120$-$250$). For the circumburst-density normalization, both LGRBs and SGRBs in the comparison sample mainly span $n_{\mathrm{env}}\approx10^{-2}$-$10^{2}$ cm$^{-3}$, but their distributions differ: LGRBs are more concentrated around $n_{\mathrm{env}}\approx 0.3$ cm$^{-3}$, whereas SGRBs are distributed more uniformly across a comparable range. For GRB~160425A, the preferred constant-density ISM fit gives $n_0\approx10^{-3}$ cm$^{-3}$, placing it at the low-density end of the sample.
Finally, model comparison favours a constant-density ISM profile for the majority of bursts, although approximately 15\% show preference for wind-like environments. Together, these results indicate that while the microphysics of shock acceleration is universal across GRB classes, their energetics, Lorentz factors  and environments reveal clear differences tied to progenitor type and central engine.

For GRB~160425A, we performed both an X-ray-only fit and a joint X-ray plus optical fit. The inferred parameters agree within the posterior uncertainties. The population comparison in Extended Data Fig.~5 uses the X-ray-only fit so that GRB~160425A is treated consistently with the comparison sample. For the multi-band afterglow fit of GRB~160425A, the ISM case gives lower values of both the  BIC  and the  DIC  than the wind case. The best-fitting ISM model has $\mathrm{BIC}=49.67$ and $\mathrm{DIC}=43.00$, compared with $\mathrm{BIC}=66.82$ and $\mathrm{DIC}=60.15$ for the wind model. The differences, $\Delta\mathrm{BIC}\approx17.15$ and $\Delta\mathrm{DIC}\approx17.15$, favour the ISM interpretation. Thus, the data are better described by a uniform circumburst medium than by a wind-like density profile.

For the preferred ISM solution, the inferred parameters are $E_{\mathrm{k,iso}}=(2.98^{+6.16}_{-1.79})\times10^{52}$ erg, $\epsilon_{e}=(6.2^{+2.5}_{-2.3})\times10^{-1}$, $\epsilon_{\rm B}=(1.2^{+5.6}_{-1.0})\times10^{-3}$, $p=2.18^{+0.04}_{-0.04}$, $\Gamma_{0}=951^{+566}_{-448}$ and $n_{0}=(1.6^{+20.0}_{-1.4})\times10^{-3}$ cm$^{-3}$. These values place GRB~160425A towards the edges of several parameter distributions of LGRBs and SGRBs (Extended Data Fig.~5). The inferred $E_{\mathrm{k,iso}}$ lies near the lower end of the LGRB population but is well above the median value for short-duration bursts ($\sim3\times10^{51}\,\mathrm{erg}$). The posterior median of $\epsilon_e$ is relatively high compared with the typical range of 0.2-0.3 for both LGRBs and SGRBs, whereas $\epsilon_B$ and $p$ lie within the distributions of the broader comparison sample. The inferred $\Gamma_0$ exceeds the median values for LGRBs ($\sim250$) and SGRBs ($\sim420$). The posterior median of $n_0$ is low, although its credible interval is broad. Within the adopted forward-shock model, these properties favour a relativistic outflow with a high initial Lorentz factor expanding into a tenuous circumburst environment.

\noindent{\bf Probability of observing two isolated GRBs within 250 s from two different galaxies at the same location }\par
\noindent  We first estimate the local coincidence probability for one specified 250-s time window and one specified BAT-scale localization region. Taking an average all-sky GRB rate of 2.5 d$^{-1}$ gives
\begin{eqnarray}
\lambda = \frac{2.5 \text{ GRBs}}{86{,}400 \text{ s}} \approx 2.89 \times 10^{-5} \text{ GRBs s}^{-1}.
\end{eqnarray}
For $T=250$ s, the Poisson probability of two GRBs in this interval is
\begin{eqnarray}
P(2;\lambda T)=\frac{(\lambda T)^2 e^{-\lambda T}}{2!}\approx 2.59\times10^{-5}.
\end{eqnarray}
Using a BAT localization radius of 1.5 arcmin, or 0.025$^\circ$, the corresponding sky fraction is
\begin{eqnarray}
f_{\rm sky}\simeq \left(\frac{0.025}{180}\pi\right)^2 \approx 7.62\times10^{-8}.
\end{eqnarray}
The local probability for two unrelated GRBs to fall within both the 250-s window and the same BAT-scale localization region is therefore
\begin{eqnarray}
P_{\rm local}\approx 2.59\times10^{-5}\times7.62\times10^{-8}\approx1.97\times10^{-12}.
\end{eqnarray}
The value above is not the false-alarm probability for the full \emph{Swift}/BAT archive. To account for the trials factor over the observing period, we estimate the expected number of unrelated GRB pairs as
\begin{eqnarray}
N_{\rm pair}\approx R_{\rm GRB}^{2} T_{\rm obs}\Delta t\, f_{\rm sky},
\end{eqnarray}
where $R_{\rm GRB}$ is the adopted GRB rate, $T_{\rm obs}$ is the \emph{Swift}/BAT observing time, $\Delta t=250$ s, and $f_{\rm sky}=7.62\times10^{-8}$ is the BAT-scale localization sky fraction used above. For $T_{\rm obs}\approx10$-20 yr, this gives a survey-level chance-coincidence probability of the order $10^{-7}$ if the \emph{Swift}/BAT trigger rate is used, and of the order $10^{-5}$ if the all-sky GRB rate of 2.5 d$^{-1}$ is adopted. We therefore quote the trials-corrected probability as $10^{-7}$-$10^{-5}$.

\noindent{\bf Probability of observing two isolated GRBs within 250 s from the same galaxy }\par
\noindent  For a GRB occurrence rate of $0.1-1$ per million years per galaxy, the rate per galaxy is
\begin{equation}
R = 3.16 \times 10^{-15} - 3.16 \times 10^{-14} \text{ GRBs s}^{-1}\text{ galaxy}^{-1}.
\end{equation}
If one GRB has already occurred in a specified galaxy, the waiting-time probability that another unrelated GRB occurs in the same galaxy within $T=250$ s is
\begin{equation}
P_{\rm local,same}=1-e^{-RT}\approx7.9\times10^{-13}-7.9\times10^{-12}.
\end{equation}
This is a local conditional probability. A trials correction over the \emph{Swift}/BAT observing period can be estimated by multiplying this value by the number of GRB triggers or candidate starting events
\begin{equation}
N_{\rm same}\approx N_{\rm trig}P_{\rm local,same}.
\end{equation}
For $N_{\rm trig}\approx10^{3}$-$10^{4}$, the corresponding probability remains of the order $10^{-9}$-$10^{-7}$. Thus, even after allowing for trials over the \emph{Swift}/BAT observing period, two unrelated bursts from the same host within minutes would still be unusual.

\noindent{\bf Uniqueness of GRB 160425A in the peculiar GRB population }\par
\noindent 
Peculiar GRBs reported so far fall into two broad observational groups based on light-curve morphology. The first comprises non-quiescent, continuous-emission outliers (top part of Table~\ref{tab:GRBtype}), including merger-like events with anomalously long $T_{90}$ (for example, GRB~060614\cite{Fynbo2006,GalYam2006,Gehrels2006}, GRB~211211A\cite{Rastinejad2022,Troja2022,Yang2022}  and GRB~230307A\cite{Levan2024,Yang2024Y}), a collapsar-like event with anomalously short $T_{90}$ (for example, GRB~200826A\cite{ZhangBB2021}), and boundary cases near the short/long classification overlap (for example, GRB~180418A\cite{Becerra2019}). Although these bursts defy duration-based classification, each constitutes a single, uninterrupted prompt-emission episode. GRB~160425A is observationally distinct: its two discrete emission episodes, separated by a   background-consistent quiescent gap, firmly place it in the  quiescent  GRB class (Fig. ~\ref{fig:lcs_PE}).

The second group comprises quiescent, multi-episode bursts (bottom part of Table~\ref{tab:GRBtype}). A number of quiescent GRBs have been reported, attracting attention owing to their prolonged quiescent periods\cite{Gotz2011,Virgili2013,HuangYY2022} or distinct properties between emission episodes\cite{Zhang2018NA,Li2019a}. Observationally, these are: LGRBs with an early precursor in which both separated components (sub-burst) are prompt-like emission and remain long-like (for example, GRB~060124\cite{Romano2006} and many other events\cite{Burlon2008,Li2026b}); SGRBs followed by a soft extended-emission tail whose post-gap component is not a conventional long-like prompt burst (for example, GRB~050709 and related events\cite{Norris2006}); double-main or double-trigger LGRBs in which both separated sub-bursts are long-like (for example, GRB~091024A\cite{Virgili2013}, GRB~110709B\cite{ZhangBB2012}  and GRB~220627A\cite{deWet2023,HuangYY2022}); LGRBs or SGRBs followed by bright X-ray flares (for example, GRB~050502B\cite{Falcone2006} and GRB~050724\cite{Dai2006}); and pre-quiescent LGRBs whose post-gap emission is more naturally interpreted as afterglow onset (for example, GRB~960530 and GRB~980125\cite{Hakkila2004}). Across all of these cases, either both separated components share a common prompt-emission class, or the post-gap component is not a second genuine prompt-emission burst. This pattern is reinforced by recent large-sample analyses showing that prompt-emission classification is internally consistent across all sub-burst episodes (precursor, main, extended  and double-main/trigger) within individual GRB events\cite{Li2026b,Li2026c,Li2026d}.

GRB~160425A satisfies none of the above precedents. Its two prompt-like episodes are separated by an ${\sim}256$-s background-consistent gap, yet their empirical classifications are opposite: $G_1$ is short-like and $G_2$ is long-like. No previously reported event combines these three properties. Continuous-emission outliers lack a separated prompt pair; previously reported quiescent bursts either preserve a common prompt-emission class across the gap or contain a post-gap component better attributed to extended emission, a flare  or afterglow onset. GRB~160425A is therefore observationally unique among known peculiar GRBs.

\section*{Data availability }

The \emph{Swift}/BAT, \emph{Swift}/XRT and \emph{Swift}/UVOT data used in this study, including the \emph{Swift}-XRT light curves used for the comparison sample, are publicly available from the UK \emph{Swift} Science Data Centre (\url{https://www.swift.ac.uk/}). The Fermi spacecraft position history and GBM daily data for 25 April 2016 are publicly available from the Fermi-GBM Daily Data archive (\url{https://heasarc.gsfc.nasa.gov/W3Browse/fermi/fermigdays.html}), and the Fermi/GBM burst-catalogue data used for the population analysis are available from the Fermi-GBM Burst Catalog (\url{https://heasarc.gsfc.nasa.gov/W3Browse/fermi/fermigbrst.html}). The optical and near-infrared follow-up measurements reported by the MPG/ESO 2.2-m telescope, CTIO/ANDICAM and MASTER-SAAO are available through the GCN circulars cited in this paper (\url{https://gcn.nasa.gov/circulars}). The reduced VLT/X-shooter spectrum is publicly available from the XSGRB archive (\url{https://sid.erda.dk/share_redirect/DBuNORk1lI/XSGRB.zip}), and the Gaia Data Release 2 astrometric data are available from the Gaia Archive (\url{https://gea.esac.esa.int/archive/}). The GROND photometry and VLT/FORS2 host-galaxy measurements obtained and analysed by the authors, together with the processed data underlying the figures and tables, are provided in the Extended Data tables and accompanying Source Data files. The comparison samples used for the quiescent gap, GRB classification, afterglow, X-ray-flare, supernova-template and host-galaxy analyses were compiled from the published catalogues and studies cited in the corresponding Methods subsections and figure legends.

\section*{Acknowledgements}

We acknowledge the use of public data from the Fermi Science Support Center, \emph{Swift} Science Data Centre and GCN circulars reported by multiple facilities. L.L. thanks F. Alamaa, N. R. Butler, K. Page, L. Izzo, P. Schady, T. Parsotan, X.-F. Wu, W.-M. Gu, W.-M. Yuan, D. Li, J.-H. Fan, E.-W. Liang, A. Lien, S. Shakeri and the ICRANet members for many helpful discussions on data analysis, GRB physics and phenomena. The computations were supported by the high-performance computing centre at Ningbo University.

\section*{Author contributions}

L.L. and Y.W. initiated the study and coordinated the scientific investigations of the event. L.L., Y.W., B.Z. and R.-G.C. were in charge of the framework of this article. L.L. organized the paper, led the data analysis and contributed to part of the physical interpretation. Y.W. discovered this particular event, assisted with the data analysis and contributed to the theoretical interpretation. L.L. and Y.W. processed and analysed the \emph{Swift}/BAT and \emph{Swift}/XRT data and the multiwavelength afterglow observations. L.L. calculated the hardness ratio, amplitude parameter, spectral lags, MVT, EHD parameter and $\varepsilon$ parameter, and performed the spectral, Amati-relation and Norris-relation analyses. B.Z., S.-N.Z., S.-R.Z., Y.-F.Y., J.-J.G. and R.-G.C. proposed the model used to explain the observed data. J.G. processed and analysed the GROND data and contributed information about the host galaxy. Y.L., Z.-P.J. and Y.-Z.F. processed and analysed the VLT data and contributed information about the host galaxy. H.-J.L. provided the source data underlying Fig. 2d and Extended Data Fig. 1d. F.R., M.K. and M.G.D. contributed to discussions of the data analysis. A.P., T.L., Y.-F.H., Z.-G.D., W.-H.L. and S.-S.X. contributed to discussions and interpretation of the results. L.L., Y.W., Y.L. and B.Z. wrote the paper with contributions from all authors. All authors reviewed, discussed and commented on the results and the paper.

\section*{Funding}

This work was supported by the National Natural Science Foundation of China (NSFC) grant numbers 12588101 (L.L. and R.-G.C.), 12233002 (Y.-F.H.), 12393813 (Y.L.), 12273113 and 12622309 (J.-J.G.), 12473012 (W.-H.L.), 12433008 (Y.-F.Y.), 12333007 (S.-N.Z.), 12494572 (T.L.) and 12225305 (Z.-P.J.). This work was financially supported by the Ningbo Science \& Technology Bureau under program grant number 2024DJRC010009 (L.L. and R.-G.C.). L.L. further acknowledges support from the Yongjiang Talent Programme (grant number 2025A-351-G). Y.L. acknowledges support from the National Key R\&D Program of China (grant number 2024YFA1611700), the Strategic Priority Research Program of the Chinese Academy of Sciences (grant number XDB0550400) and the China Manned Space Program (grant number CMS-CSST-2025-A17). Y.-F.H. was also supported by the National Key R\&D Program of China (grant number 2021YFA0718500) and by the Xinjiang Tianchi Program. M.G.D. acknowledges support from the JSPS Grant-in-Aid for Scientific Research (KAKENHI) (A) (grant number JP25H00675).

\section*{Competing interests}

The authors declare no competing interests.

\section*{Additional information}

Extended data is available for this paper at https://doi.org/10.1038/s41550-026-03007-0. Supplementary information The online version contains supplementary material available at https://doi.org/10.1038/s41550-026-03007-0.Correspondence and requests for materials should be addressed to Liang Li, Yu Wang, Bing Zhang or Rong-Gen Cai. Peer review information Nature Astronomy thanks Anya Nugent, Genevieve Schroeder and the other, anonymous, reviewer(s) for their contribution to the peer review of this work. Peer reviewer reports are available. Reprints and permissions information is available at www.nature.com/reprints.

\noindent Publisher's note Springer Nature remains neutral with regard to jurisdictional claims in published maps and institutional affiliations. Springer Nature or its licensor (e.g. a society or other partner) holds exclusive rights to this article under a publishing agreement with the author(s) or other rightsholder(s); author self-archiving of the accepted manuscript version of this article is solely governed by the terms of such publishing agreement and applicable law. \copyright{} The Author(s), under exclusive licence to Springer Nature Limited 2026

\setcounter{figure}{0}    
\setcounter{section}{0}
\setcounter{table}{0}

\captionsetup[table]{name={\bf Extended Data Table}}
\captionsetup[figure]{name={\bf Extended Data Fig.}}

\clearpage
\begin{figure}[ht!]
\centering
\includegraphics[width=\textwidth,height=0.62\textheight,keepaspectratio]{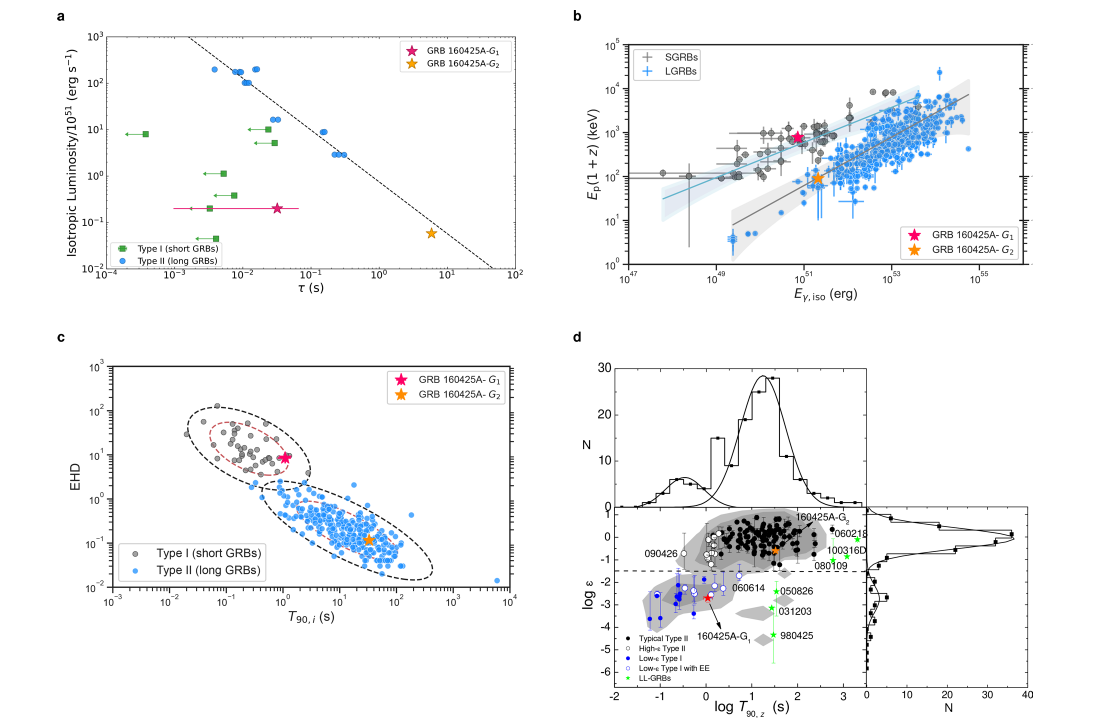}
\caption{{\bf Alternative prompt-emission classification diagnostics for GRB 160425A.} {\bf a-d} illustrate additional classification diagnostics using the Norris relation ({\bf a}), the Amati relation ({\bf b}), the duration/EHD diagram ({\bf c}), and the duration/$\epsilon$ diagram ({\bf d}). In {\bf a}, Type I and Type II GRBs are represented by green squares and blue circles, respectively; the black dashed line shows the best-fitting Norris relation, and left-pointing arrows indicate upper limits. In {\bf b}, short- and long-duration GRBs are shown in grey and blue, respectively; solid lines show the best-fitting relations, and the shaded bands indicate the corresponding 1$\sigma$ and 2$\sigma$ regions. In {\bf c}, Type I and Type II GRBs are shown in grey and blue, respectively, with dashed contours indicating their empirical distributions. In {\bf d}, typical and high-$\epsilon$ Type II GRBs are shown by filled and open black symbols, respectively; low-$\epsilon$ Type I GRBs without and with extended emission are shown by filled and open blue symbols, respectively; and low-luminosity GRBs are shown by green stars. The grey shaded regions indicate the loci of the different populations, the horizontal dashed line marks the division at $\epsilon$ = 0.03, and the marginal distributions are shown along the top and right axes, with solid curves indicating Gaussian fits. In all panels, $G_1$ and $G_2$ of GRB 160425A are highlighted by red and orange stars, respectively. Symbols denote the reported best-fitting values for individual GRBs, and the error bars in {\bf a}, {\bf b} and {\bf d} represent the corresponding 1$\sigma$ measurement uncertainties. Durations in {\bf c} and {\bf d} are given in the source frame.}
\end{figure}

\clearpage
\begin{figure}[ht!]
\centering
\includegraphics[width=\textwidth,height=0.62\textheight,keepaspectratio]{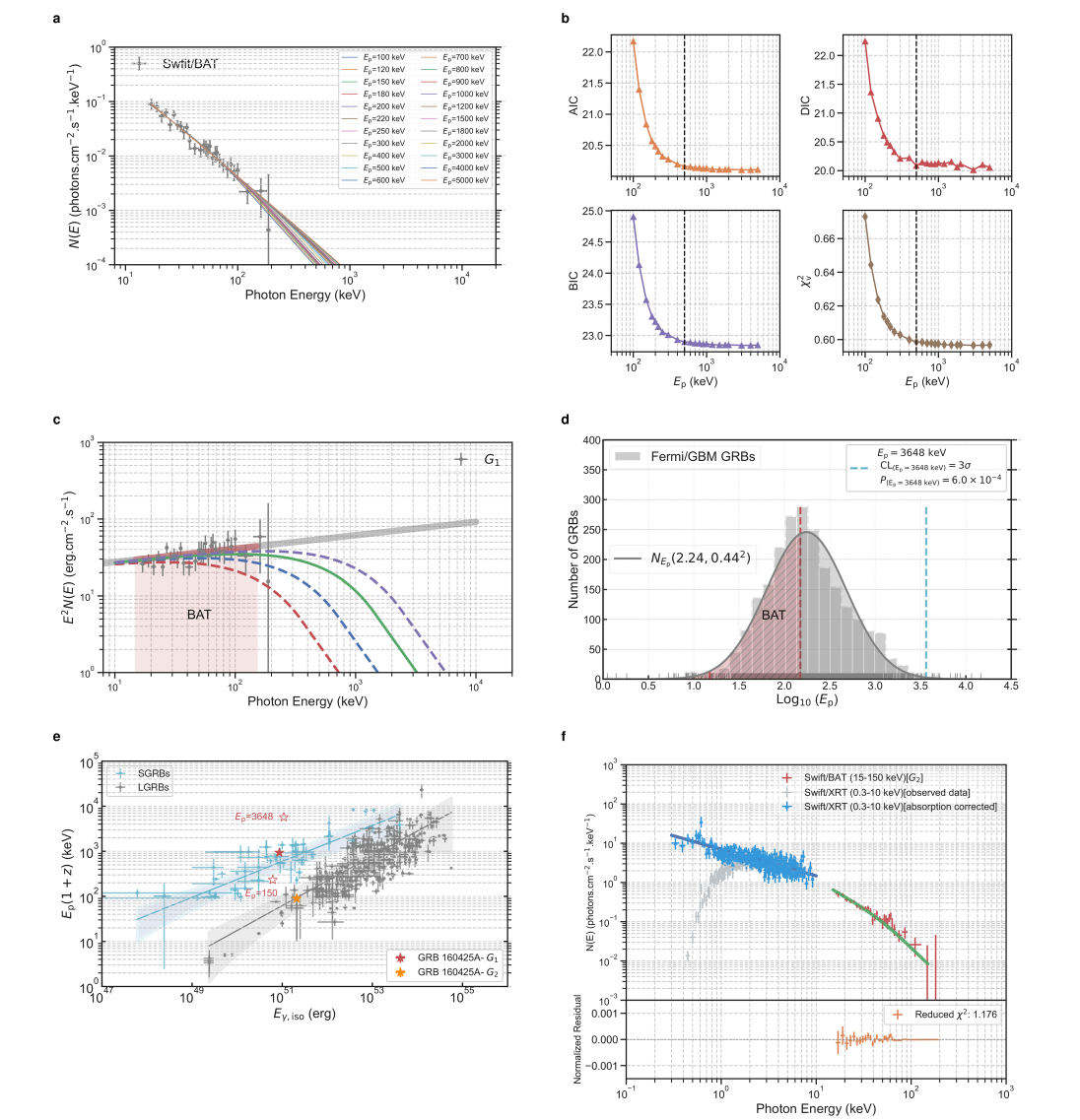}
\caption{{\bf Spectral fit results for $G_1$ and $G_2$.} {\bf a-c}, The estimated $E_{\rm p}$ value of $G_1$ using spectral-shape curvature extrapolation method. {\bf a}, The BAT spectra data of $G_1$ and their best fit using the different Band models with varying $E_{\rm p}$ values. The symbols show the measured spectral values; vertical error bars represent the corresponding 1$\sigma$ statistical uncertainties, whereas horizontal bars indicate the widths of the energy bins. {\bf b}, The corresponding statistical criteria as a function of $E_{\rm p}$, with the vertical black dashed line indicating the ``bending point''. {\bf c}, The corresponding Band function with different $E_{\rm p}$ values, indicated by differently colored dashed and solid lines, with the solid line corresponding to the ``bending point''. The vertical error bars on the spectral data represent 1$\sigma$ statistical uncertainties, whereas the horizontal bars indicate the widths of the energy bins. The red shaded region marks the \emph{Swift}-BAT energy range and does not represent an uncertainty interval. {\bf d}, The upper limit of $E_{\rm p}$ was estimated using a complete GBM GRB sample. {\bf e}, The Amati $E_{\gamma,\rm iso}-E_{\rm p,z}$ correlation diagram for short-duration (cyan data points) and long-duration (gray) as well as their best model fittings (solid lines) and 2$\sigma$ regions (shaded areas). The symbols show the reported values for individual GRBs, and the horizontal and vertical error bars represent the corresponding 1$\sigma$ measurement uncertainties. The two individual bursts of GRB are highlighted by red and orange stars, respectively. {\bf f}, The BAT-XRT spectra of $G_2$ and the best fits using the Band model. The symbols show the measured spectral values; vertical error bars represent 1$\sigma$ statistical uncertainties, horizontal bars indicate the widths of the energy bins, and the error bars in the residual panel represent the propagated 1$\sigma$ statistical uncertainties.}
\end{figure}

\clearpage
\begin{figure}[ht!]
\centering
\includegraphics[width=\textwidth,height=0.62\textheight,keepaspectratio]{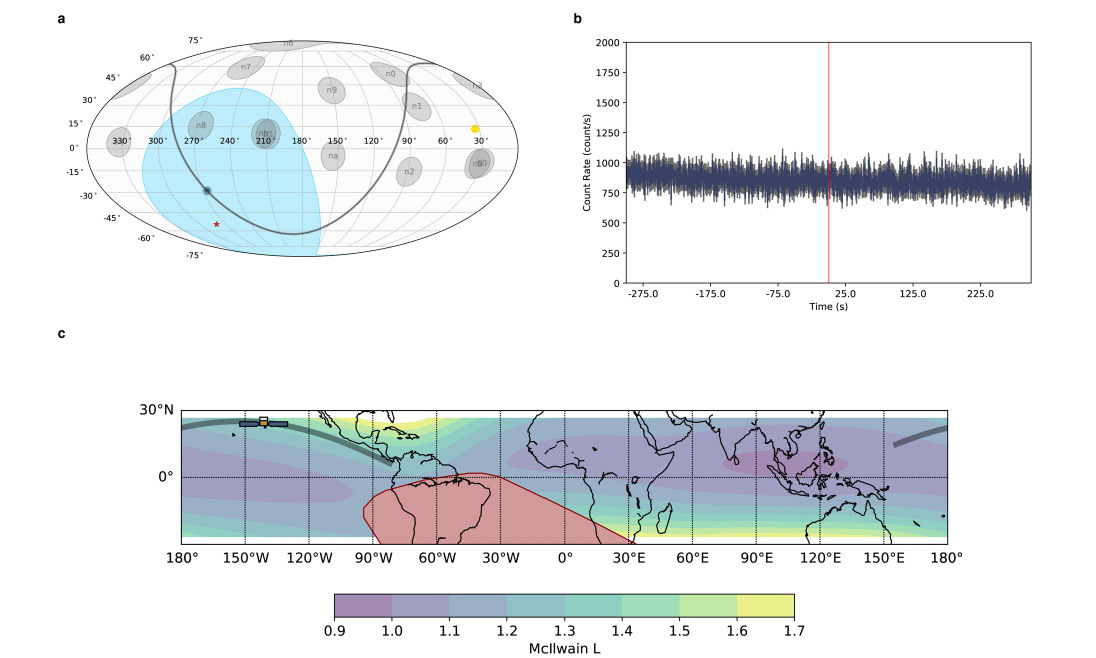}
\caption{{\bf Untriggered Fermi-GBM signal search and data analysis.} {\bf a}, Position history (POSHIST) data of the Fermi satellite from the Fermi GBM Daily Data for April 25, 2016. The red pentagram marks the location of GRB 160425A, while the blue region represents the Earth. {\bf b}, The all-sky light curve observed by Fermi, shows no clear signal within 5 minutes before and after the trigger time. {\bf c}, Trajectory of the Fermi satellite during the burst of GRB 160425A.}
\end{figure}

\clearpage
\begin{figure}[ht!]
\centering
\includegraphics[width=\textwidth,height=0.62\textheight,keepaspectratio]{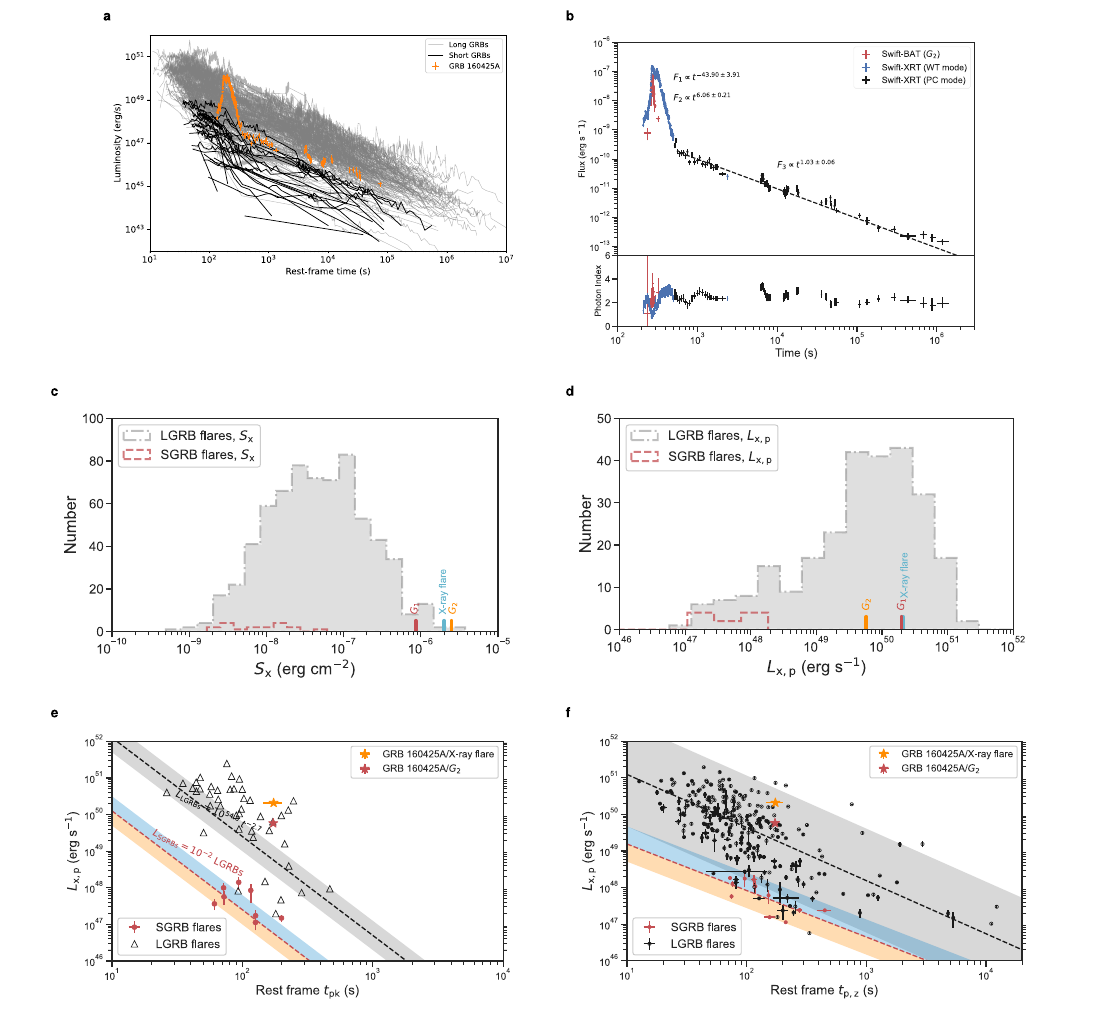}
\caption{{\bf \emph{Swift}-XRT afterglow emissions of GRB 160425A.} {\bf a}, The XRT luminosity as a function of rest-frame time for short-duration (black) and long-duration (gray) GRBs, based on a GRB sample detected from 2004 to 2015, with GRB 160425A indicated by orange data points. {\bf b}, X-ray light curve and photon index evolution of GRB 160425A, as well as the best-fitting of the X-ray flare and normal decay phase. {\bf c-d}, The distributions of energy fluence ($S_{\rm X}$) and peak luminosity ($L_{\rm x,p}$) for XRT flares in short-duration (red) and long-duration (grey) GRBs, based on a comprehensive catalog of short-duration GRB flares and long-duration GRB flares\cite{Shi2022Univ}. The position of $G_1$ and $G_2$ of GRB 160425A and its X-ray flare in the distributions is marked by red, orange, and cyan vertical cylinders, respectively. These coloured vertical markers indicate the measured central values only; their heights are used for identification and do not represent measurement uncertainties. {\bf e}, The luminosity of XRT flares as a function of rest-frame time for short-duration (red) and long-duration (grey) GRBs as well as their best model fittings (dashed lines) and 2$\sigma$ regions (shaded areas), derived from a catalog of short-duration GRB flares\cite{Margutti2011} and long-duration GRB flares\cite{Chincarini2010}. The symbols represent measurements of individual XRT flares, and the horizontal and vertical error bars, where shown, represent the corresponding 1$\sigma$ measurement uncertainties. The X-ray flare and $G_2$ of GRB 160425A are highlighted by orange (the X-ray flare) and red ($G_2$) stars, respectively. {\bf f}, The same as {\bf e}, but for a different XRT flare catalog\cite{Shi2022Univ}.}
\end{figure}

\clearpage
\begin{figure}[ht!]
\centering
\includegraphics[width=\textwidth,height=0.62\textheight,keepaspectratio]{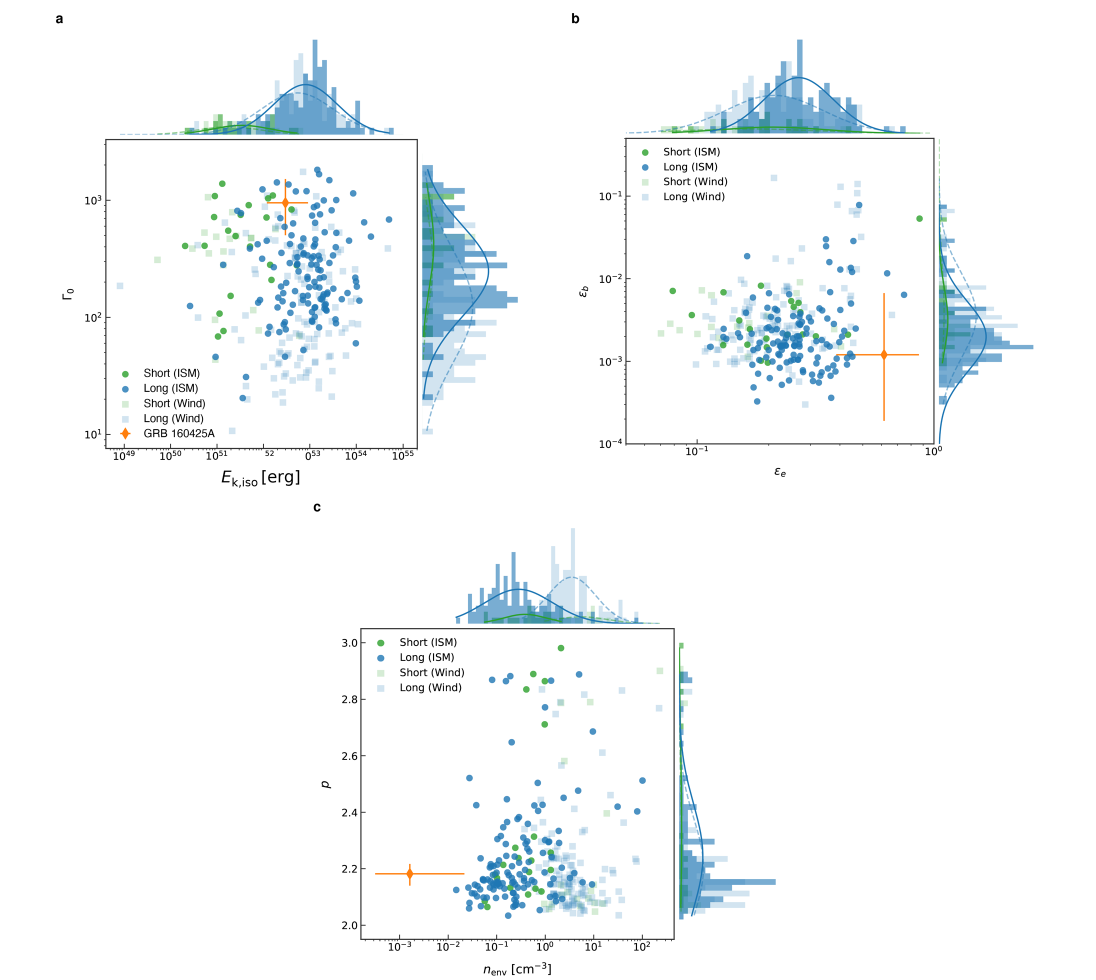}
\caption{{\bf Synchrotron afterglow modelling of GRB 160425A in comparison with \emph{Swift} GRBs.} {\bf a}, The relationship between the isotropic equivalent energy, $E_{\rm k,iso}$, and the initial Lorentz factor, $\Gamma_0$. The data is distinguished by burst class (Short-duration and long-duration GRBs) and by the assumed circumstellar medium model (ISM and Wind). The central panel displays a log-log scatter plot of $E_{\rm k,iso}$ versus $\Gamma_0$. The top and right panels show the corresponding distributions of these parameters as histograms. The values for GRB 160425A are shown with an error bar. {\bf b}, The median values of the microphysical parameters, $\epsilon_e$ (electron energy fraction) and $\epsilon_b$ (magnetic field energy fraction). The data points are differentiated by burst class (Short-duration and long-duration GRBs) and circumstellar medium model (ISM and Wind). The central panel is a log-log scatter plot of $\epsilon_e$ versus $\epsilon_b$. The top and right panels show the distributions of $\epsilon_e$ and $\epsilon_b$, respectively, as histograms. Overlaid on the histograms are Gaussian fits to the data in logarithmic space. The specific values for GRB 160425A are highlighted with an error bar. {\bf c}, The circumstellar medium density, $n_{\rm env}$ (in cm$^{-3}$), and the electron energy distribution index, p. The data points are categorized by the GRB class (Short and Long) and the specific model used for the circumstellar medium (ISM or Wind). The central panel is a log-log scatter plot of $n_{\rm env}$ against p. The distributions for each parameter are displayed as histograms on the top (p) and right ($n_{\rm env}$) axes. The data point for GRB 160425A is highlighted with its associated error bar. In all panels, the orange diamonds show the posterior median values inferred for GRB 160425A; the horizontal and vertical error bars extend from the 16th to the 84th percentiles of the marginalized posterior distributions, corresponding to 68\% credible intervals. The marginal histograms show the distributions of posterior median values for the comparison samples, and the overlaid curves show fits to these distributions rather than uncertainty bands.}
\end{figure}

\clearpage
\begin{figure}[ht!]
\centering
\includegraphics[width=\textwidth,height=0.62\textheight,keepaspectratio]{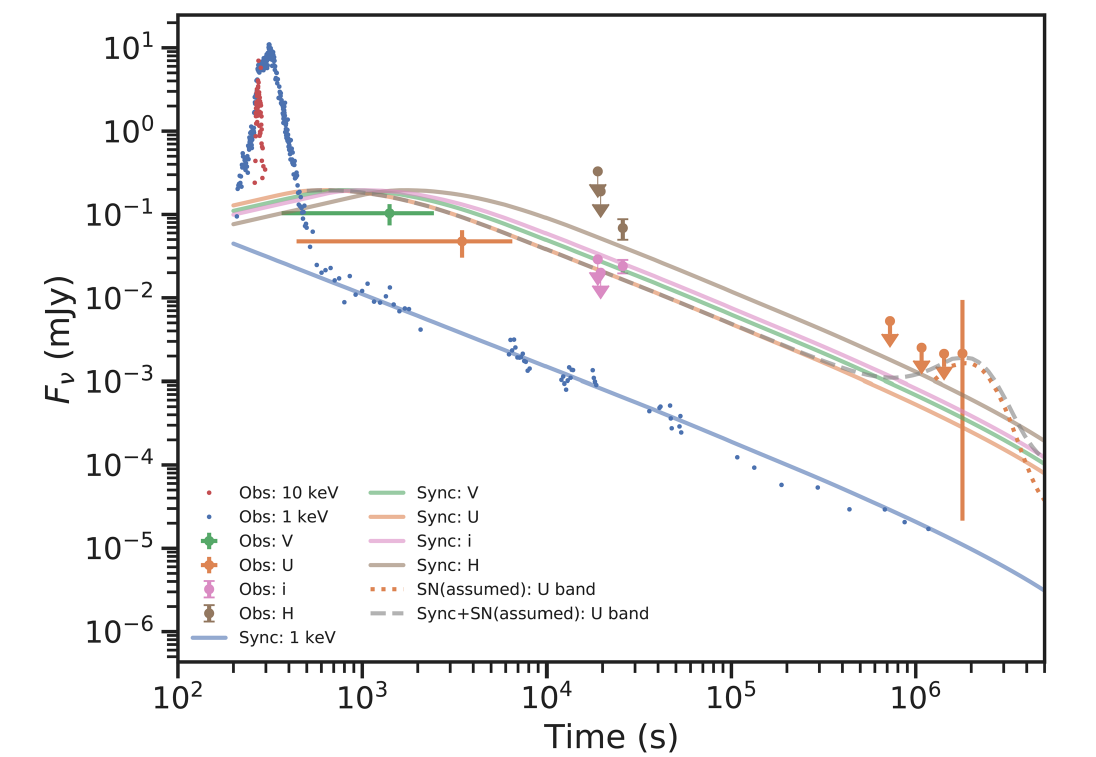}
\caption{{\bf Multi-wavelength observations and modelling of GRB 160425A.} The multi-wavelength data (X-ray observed by \emph{Swift}-BAT and XRT, U and V bands observed by \emph{Swift}-UVOT, i and H bands observed by GROND and ESO-2.2m MPG) including detections and upper limits, are represented by solid circles and downward arrows, respectively. The best-fit afterglow by the synchrotron emission is represented by dashed curves, the SN component by a dotted curve. Optical data cannot determine the presence of a supernova component. To estimate the maximum luminosity of a potential supernova, we added an orange dotted line (one magnitude fainter than SN 1998bw) and a gray dashed line to represent the hypothetical light curve of the supernova in the U band and the total light curve after incorporating synchrotron radiation, respectively.}
\end{figure}

\clearpage
\begin{figure}[ht!]
\centering
\includegraphics[width=\textwidth,height=0.62\textheight,keepaspectratio]{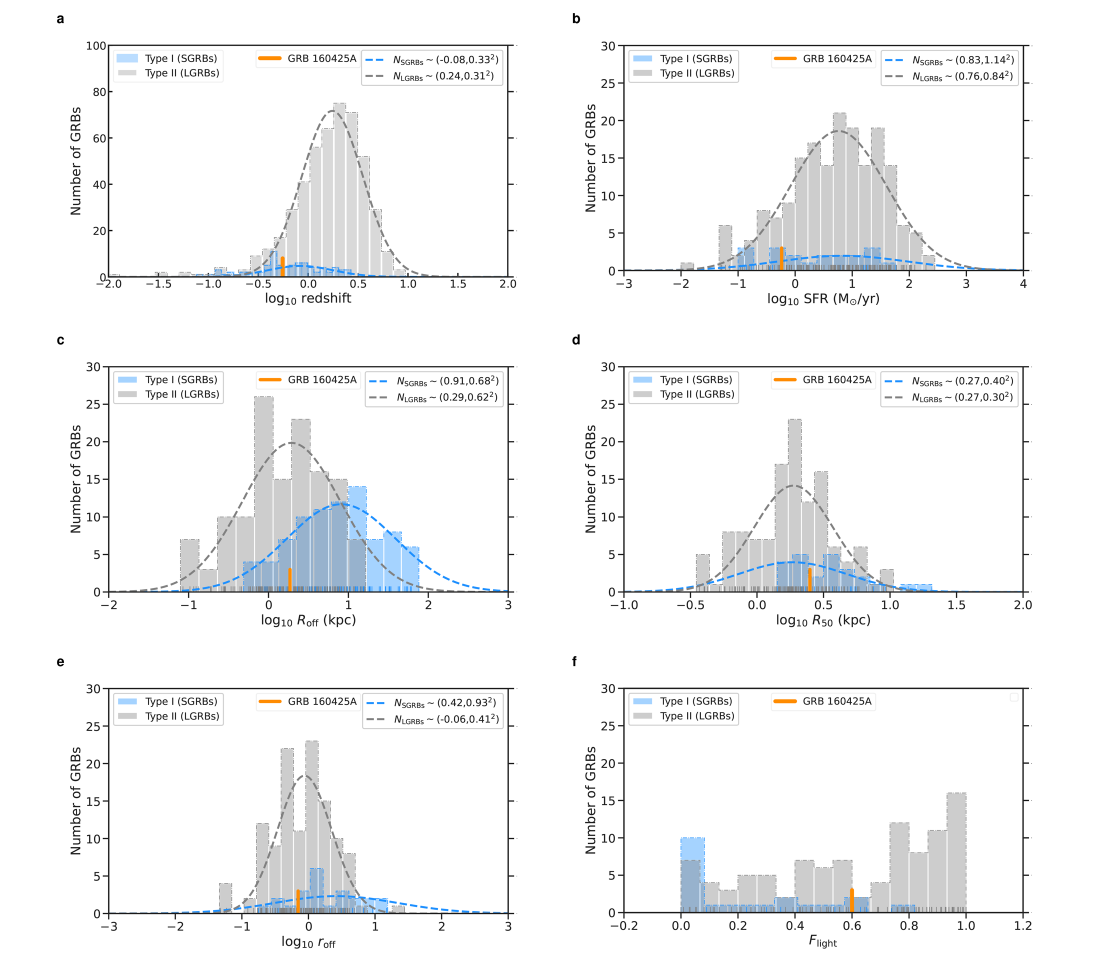}
\caption{{\bf Host properties of GRB 160425A.} {\bf a}, Distributions of redshift. {\bf b}, Distributions of star formation rate (SFR) of the host galaxies. {\bf c}, Distributions of offset $R_{\rm off}$. {\bf d}, Distributions of half light radius $R_{50}$. {\bf e}, Distributions of normalized offset ($r=R_{\rm off}/R_{50}$). {\bf f}, Distributions of cumulative light fraction $F_{\rm light}$. In all panels ({\bf a}-{\bf f}), blue and gray histograms represent Type I and Type II GRBs, respectively, with the orange vertical cylinder bar indicating the position of GRB 160425A.}
\end{figure}

\clearpage
\begin{table}
\setlength{\tabcolsep}{0.50em}
\renewcommand\arraystretch{1.2}
\caption{{\bf Ultraviolet, Optical and Near-infrared Observations of GRB 160425A}\label{tab:Optobs}}
\centering
\begingroup\fontsize{6.5}{7.8}\selectfont\renewcommand\arraystretch{1.0}
\begin{tabular}{ccccc}
\hline
$T-T_{0}$ (day)&AB Magnitude&Filter&Telescope&Ref.\\
\hline
0.005$^{+0.001}_{-0.001}$&$>$18.93&V&{\it Swift}-UVOT&\cite{2016GCN.19357....1S}\\
0.007$^{+0.001}_{-0.001}$&$>$18.36&W2&{\it Swift}-UVOT&\cite{2016GCN.19357....1S}\\
0.007$^{+0.002}_{-0.002}$&$>$18.87&M2&{\it Swift}-UVOT&\cite{2016GCN.19357....1S}\\
0.007$^{+0.002}_{-0.002}$&$>$19.43&U&{\it Swift}-UVOT&\cite{2016GCN.19357....1S}\\
0.007$^{+0.002}_{-0.002}$&$>$20.06&W1&{\it Swift}-UVOT&\cite{2016GCN.19357....1S}\\
0.008$^{+0.002}_{-0.002}$&$>$20.92&B&{\it Swift}-UVOT&\cite{2016GCN.19357....1S}\\
0.012$^{+0.004}_{-0.004}$&$>$19.87&V&{\it Swift}-UVOT&\cite{2016GCN.19357....1S}\\
0.013$\pm$0.008&20.16$\pm$0.32&b&{\it Swift}-UVOT&\cite{2016GCN.19357....1S}\\
0.016$\pm$0.012&19.22$\pm$0.31&v&{\it Swift}-UVOT&\cite{2016GCN.19357....1S}\\
0.017$^{+0.005}_{-0.005}$&$>$19.26&W2&{\it Swift}-UVOT&\cite{2016GCN.19357....1S}\\
0.017$^{+0.004}_{-0.004}$&$>$20.94&B&{\it Swift}-UVOT&\cite{2016GCN.19357....1S}\\
0.018$^{+0.005}_{-0.005}$&$>$20.71&W1&{\it Swift}-UVOT&\cite{2016GCN.19357....1S}\\
0.019$^{+0.006}_{-0.006}$&$>$19.52&U&{\it Swift}-UVOT&\cite{2016GCN.19357....1S}\\
0.019$^{+0.006}_{-0.006}$&$>$20.31&M2&{\it Swift}-UVOT&\cite{2016GCN.19357....1S}\\
0.023$^{+0.005}_{-0.005}$&$>$19.28&V&{\it Swift}-UVOT&\cite{2016GCN.19357....1S}\\
0.025$^{+0.002}_{-0.002}$&$>$20.55&B&{\it Swift}-UVOT&\cite{2016GCN.19357....1S}\\
0.026$^{+0.002}_{-0.002}$&$>$18.55&W2&{\it Swift}-UVOT&\cite{2016GCN.19357....1S}\\
0.026$^{+0.001}_{-0.001}$&$>$18.94&W1&{\it Swift}-UVOT&\cite{2016GCN.19357....1S}\\
0.027$^{+0.001}_{-0.001}$&$>$17.86&U&{\it Swift}-UVOT&\cite{2016GCN.19357....1S}\\
0.028$^{+0.001}_{-0.001}$&$>$17.89&M2&{\it Swift}-UVOT&\cite{2016GCN.19357....1S}\\
0.080$^{+0.009}_{-0.010}$&$>$22.12&W1&{\it Swift}-UVOT&\cite{2016GCN.19357....1S}\\
0.083$^{+0.010}_{-0.010}$&$>$20.10&U&{\it Swift}-UVOT&\cite{2016GCN.19357....1S}\\
0.085$^{+0.011}_{-0.009}$&$>$21.28&B&{\it Swift}-UVOT&\cite{2016GCN.19357....1S}\\
0.087$^{+0.001}_{-0.001}$&$>$21.13&M2&{\it Swift}-UVOT&\cite{2016GCN.19357....1S}\\
0.119$^{+0.039}_{-0.040}$&$>$20.92&W2&{\it Swift}-UVOT&\cite{2016GCN.19357....1S}\\
0.122$^{+0.039}_{-0.039}$&$>$21.07&V&{\it Swift}-UVOT&\cite{2016GCN.19357....1S}\\
0.211$^{+0.006}_{-0.009}$&$>$21.78&B&{\it Swift}-UVOT&\cite{2016GCN.19357....1S}\\
0.227&$>$19.5&I&CTIO-1.3m ANDICAM&\cite{2016GCN.19354....1C}\\
0.227&$>$16.6&J&CTIO-1.3m ANDICAM&\cite{2016GCN.19354....1C}\\
0.227&$>$15.7&K&CTIO-1.3m ANDICAM&\cite{2016GCN.19354....1C}\\
0.298$\pm$0.008&22.2$\pm$0.3&g&ESO-2.2m MPG&\cite{2016GCN.19349....1B}\\
0.298$\pm$0.008&21.1$\pm$0.2&r&ESO-2.2m MPG&\cite{2016GCN.19349....1B}\\
0.298$\pm$0.008&20.7$\pm$0.2&i&ESO-2.2m MPG&\cite{2016GCN.19349....1B}\\
0.298$\pm$0.008&20.5$\pm$0.1&z&ESO-2.2m MPG&\cite{2016GCN.19349....1B}\\
0.298$\pm$0.008&19.7$\pm$0.2&J&ESO-2.2m MPG&\cite{2016GCN.19349....1B}\\
0.298$\pm$0.008&19.4$\pm$0.3&H&ESO-2.2m MPG&\cite{2016GCN.19349....1B}\\
0.298$\pm$0.008&$>$18.3&K&ESO-2.2m MPG&\cite{2016GCN.19349....1B}\\
0.377&$>$19.9&I&CTIO-1.3m ANDICAM&\cite{2016GCN.19354....1C}\\
0.377&$>$17.9&J&CTIO-1.3m ANDICAM&\cite{2016GCN.19354....1C}\\
0.377&$>$17.1&K&CTIO-1.3m ANDICAM&\cite{2016GCN.19354....1C}\\
0.04$\pm$0.035&20.07$\pm$0.26&u&{\it Swift}-UVOT&\cite{2016GCN.19357....1S}\\
0.047$\pm$0.042&20.24$\pm$0.30&w1&{\it Swift}-UVOT&\cite{2016GCN.19357....1S}\\
0.082$\pm$0.076&$>$20.8&w2&{\it Swift}-UVOT&\cite{2016GCN.19357....1S}\\
0.046$\pm$0.041&$>$19.7&m2&{\it Swift}-UVOT&\cite{2016GCN.19357....1S}\\
7.384$^{+0.287}_{-0.249}$&$>$21.74&W1&{\it Swift}-UVOT&\cite{2016GCN.19357....1S}\\
8.389$^{+0.293}_{-0.314}$&$>$21.75&U&{\it Swift}-UVOT&\cite{2016GCN.19357....1S}\\
9.714$^{+0.130}_{-0.129}$&21.74$^{+0.37}_{-0.29}$&W2&{\it Swift}-UVOT&\cite{2016GCN.19357....1S}\\
10.613&$>$20.74&M2&{\it Swift}-UVOT&\cite{2016GCN.19357....1S}\\
11.427&$>$21.74&W1&{\it Swift}-UVOT&\cite{2016GCN.19357....1S}\\
12.448$^{+0.329}_{-0.320}$&$>$22.04&U&{\it Swift}-UVOT&\cite{2016GCN.19357....1S}\\
13.543$^{+0.430}_{-0.417}$&21.43$^{+0.33}_{-0.25}$&W2&{\it Swift}-UVOT&\cite{2016GCN.19357....1S}\\
14.499$^{+0.383}_{-0.397}$&21.71$^{+0.44}_{-0.33}$&M2&{\it Swift}-UVOT&\cite{2016GCN.19357....1S}\\
16.485&$>$22.08&U&{\it Swift}-UVOT&\cite{2016GCN.19357....1S}\\
18.576&21.80$^{+0.44}_{-0.31}$&M2&{\it Swift}-UVOT&\cite{2016GCN.19357....1S}\\
20.743&21.89$^{+0.36}_{-0.27}$&U&{\it Swift}-UVOT&\cite{2016GCN.19357....1S}\\
\hline
\end{tabular}
\endgroup
\begin{flushleft}
Ultraviolet, optical and near-infrared follow-up observations of GRB 160425A used in the multi-wavelength afterglow analysis. The measurements were obtained from \emph{Swift}/UVOT observations and the ground-based follow-up reports cited in Methods.
\end{flushleft}
\end{table}

\clearpage
\begin{table}
\setlength{\tabcolsep}{0.50em}
\renewcommand\arraystretch{1.2}
\caption{\bf Optical and Near-infrared GROND Observations of GRB 160425A\label{tab:OptobsGROND}}
\centering
\begin{tabular}{ccccc}
\hline
$T-T_{0}$ (day)&AB Magnitude&Filter&Telescope&Ref.\\
\hline
0.218&$>$20.7&g&GROND&this paper\\
0.218&$>$20.9&r&GROND&this paper\\
0.218&$>$20.5&i&GROND&this paper\\
0.218&$>$20.5&z&GROND&this paper\\
0.218&$>$18.4&J&GROND&this paper\\
0.218&$>$17.7&H&GROND&this paper\\
0.218&$>$17.3&K&GROND&this paper\\
0.226&$>$20.9&g&GROND&this paper\\
0.226&$>$21.2&r&GROND&this paper\\
0.226&$>$20.9&i&GROND&this paper\\
0.226&$>$20.8&z&GROND&this paper\\
0.226&$>$18.6&J&GROND&this paper\\
0.226&$>$18.3&H&GROND&this paper\\
0.226&$>$17.4&K&GROND&this paper\\
\hline
\end{tabular}
\begin{flushleft}
Optical and near-infrared follow-up observations of GRB 160425A obtained with GROND on the MPG/ESO 2.2-m telescope at La Silla Observatory. These measurements are used in the multi-wavelength afterglow analysis described in Methods.
\end{flushleft}
\end{table}

\clearpage
\begin{table}
\setlength{\tabcolsep}{0.10em}
\renewcommand\arraystretch{1.2}
\caption{\bf GRB 160425A in the Statistical Results of Parameter Distributions of Host Properties\label{tab:host}}
\centering
\begin{tabular}{c|c|c|c|c|c|c|c|c|c}
\hline
Name&
\multicolumn{2}{c|}{SGRBs}&
\multicolumn{2}{c|}{LGRBs}&
\multicolumn{4}{c|}{GRB160425A}&
Classfication\\
\hline
&Number&$N_{\rm SGRB}(\mu,\sigma^{2})$&Number&$N_{\rm LGRB}(\mu,\sigma^{2})$&Confidence level&$P_{\rm SGRB}$&Confidence level&$P_{\rm LGRB}$&Short vs. Long\\
\hline
$z$&78&(-0.08,0.33$^{2}$)&484&$(0.24,0.31^{2})$&0.54$\sigma$&0.29&1.63$\sigma$&0.05&Short\\
SFR&19&$(0.83,1.14^{2})$&183&$(0.76,0.84^{2})$&0.93$\sigma$&0.18&1.18$\sigma$&0.12&Short\\
$R_{\rm off}$&83&$(0.91,0.68^{2})$&132&$(0.29,0.62^{2})$&0.94$\sigma$&0.17&0.03$\sigma$&0.49&Long\\
$R_{\rm 50}$&24&$(0.27,0.40^{2})$&124&$(0.27,0.30^{2})$&0.32$\sigma$&0.63&0.41$\sigma$&0.66&Short\\
$r_{\rm off}$&23&$(0.42,0.93^{2})$&118&(-0.06,0.41$^{2}$)&0.62$\sigma$&0.27&0.24$\sigma$&0.41&Long\\
$F_{\rm light}$&20&Unconstrained&99&Unconstrained&...&...&...&...&Long\\
\hline
\end{tabular}

\begin{flushleft}
Comparison of the host-galaxy properties of GRB 160425A with those of short- and long-duration GRBs. The analysis considers redshift, star formation rate, projected physical offset, half-light radius, normalized offset and cumulative light fraction. Only secure measurements are included; upper and lower limits are excluded. The statistical procedures used to compare GRB 160425A with the two populations are described in Methods.	
\end{flushleft}

\end{table}

\end{document}